\documentclass[aps,prev,twocolumn,preprintnumbers,floatfix,nofootinbib]{revtex4-1}

\usepackage{graphicx}
\usepackage[ngerman,english]{babel}
\usepackage{bm}
\usepackage{times}
\usepackage{slashed}
\usepackage{mathtools}
\usepackage{color}
\usepackage{epsfig}
\usepackage{dcolumn}
\usepackage{textcomp}
\usepackage{color}

\begin{document} 

\title{\boldmath The Dark $L_\mu - L_\tau$ Rises via Kinetic Mixing}

\author{Giorgio Arcadi$^1$}
\author{Thomas Hugle$^1$}
\author{Farinaldo S. Queiroz$^{2}$}

\email{arcadi@mpi-hd.mpg.de}
\email{thomas.hugle@mpi-hd.mpg.de}
\email{farinaldo.queiroz@iip.ufrn.br}

\affiliation{$^1$Max-Planck-Institut f\"ur Kernphysik (MPIK),
Saupfercheckweg 1, 69117 Heidelberg, Germany}
\affiliation{$^2$International Institute of Physics, Federal University of Rio Grande do Norte,
Campus Universitario, Lagoa Nova, Natal-RN 59078-970, Brazil}

\begin{abstract}
We investigate the dark matter phenomenology of a Dirac fermion together with kinetic mixing in the context of an $L_\mu - L_\tau$ model. We analyze which part of the parameter space can provide a viable dark matter candidate and explain the B-decay anomaly, while obeying current data. Although the allowed region of parameter space, satisfying these requirements, is still large at the moment, future direct detection experiments like XENONnT and DARWIN will, in case of null results, significantly strengthen the limits and push the model to a corner of the parameter space.
\end{abstract}

\maketitle
\flushbottom

%%%%%%%%%%%%%%%%%%%%%%%%%%%%%%%%%%%%%%%%%%%%%%%%%%%%%%%%%%%%%%%%%%%%%%%%%%%%%%%%%%
\section{Introduction}
\label{sec_introduction}
%%%%%%%%%%%%%%%%%%%%%%%%%%%%%%%%%%%%%%%%%%%%%%%%%%%%%%%%%%%%%%%%%%%%%%%%%%%%%%%%%%

Precise measurement of the cosmic microwave background (CMB) radiation has greatly impacted our understanding of the universe. In particular, it has shown that 27\% of the energy budget of our universe is comprised of dark matter (DM) \cite{Ade:2015xua} whose nature remains a mystery \cite{Queiroz:2016sxf,Catena:2017xqq,Kavanagh:2017hcl}. The success of the CMB and Big Bang Nucleosynthesis suggests thermal equilibrium as a fundamental guiding principle. Therefore, it is plausible to assume that dark matter particles were also in thermal equilibrium with the Standard Model particles in the early universe. This key assumption implies, for the DM, a thermal annihilation cross section at the weak scale which has been intensively probed via indirect detection experiments \cite{Ackermann:2015zua,Wood:2015ofa,Abdallah:2016ygi,Ahnen:2016qkx,Archambault:2017wyh}. In this work, we assess this thermal dark matter production in the context of the $L_\mu - L_\tau$ symmetry.\\

Extensions of the Standard Model (SM) involving the difference of the lepton numbers such as the $L_\mu-L_\tau$ gauge group are automatically anomaly free \cite{Foot:1990mn,Foot:1994vd}. Since this symmetry does not act on the the first generation of leptons, the model can potentially explain the anomaly  observed concerning the muon anomalous magnetic moment \cite{Lindner:2016bgg,Heeck:2011wj} without being subject to the strong limits from the Large Electron-Positron (LEP) collider \cite{Freitas:2014pua}. Many other phenomenological studies have been carried out in the context of the $L_\mu-L_\tau$ symmetry focusing on flavor anomalies \cite{Altmannshofer:2015mqa,Omura:2015xcg,Altmannshofer:2016oaq,Ibe:2016dir,Baek:2017sew,Ko:2017yrd,Chen:2017usq,Chen:2017cic,Gninenko:2018tlp,Bian:2017rpg,Bian:2017xzg}, neutrino \cite{Gupta:2013it,Chen:2017gvf,Dev:2017fdz,Nath:2018hjx} or collider physics \cite{Elahi:2015vzh,Kaneta:2016uyt,Araki:2017wyg,Nomura:2018yej,Kim:2015fpa,Kohda:2018xbc,Hou:2018npi}. In this work, we are mainly interested in the dark matter phenomenology. In \cite{Biswas:2016yjr} feebly interacting particles were investigated as dark matter, while in \cite{Baek:2008nz,Altmannshofer:2016jzy} Weakly Interacting Massive Particles (WIMPs) were the candidates. In the latter, the authors discussed a model where there was no $Z-Z'$ kinetic mixing. In \cite{Baek:2015fea,Biswas:2016yan,Patra:2016shz} different incarnations of the $L_\mu-L_\tau$ symmetry were addressed. \\

Our work differs from previous studies because we introduce a kinetic mixing parameter which is expected to be present since it is gauge invariant and there is no reason for it to be neglected. Even if one sets it to zero at tree level it can be generated at 1-loop \cite{Holdom:1985ag,Carone:1995pu,Kahlhoefer:2015bea}. This crucial point leads to a different dark matter phenomenology because now the dark matter candidate can annihilate into all SM fermions and gauge bosons and scatter off nuclei at tree-level. \\

We take the opportunity to assess whether this model can also address the anomaly observed in the LHCb data concerning the rare decay of B-mesons \cite{Descotes-Genon:2013wba} ($B \rightarrow K^{\ast} \mu^+ \mu^-$). This anomaly observed in LHCb data by \cite{Descotes-Genon:2013wba} has been confirmed by several independent studies \cite{Altmannshofer:2013foa,Beaujean:2013soa,Hurth:2013ssa} including a new one from LHCb \cite{Aaij:2015oid} which also hints at new signatures such as lepton flavor non-universality \cite{Aaij:2014ora}. These findings are furthermore supported by global analyses \cite{Altmannshofer:2011gn,Altmannshofer:2012az,Altmannshofer:2014rta,Descotes-Genon:2015uva,Neshatpour:2017qvi,Capdevila:2017bsm,Altmannshofer:2017fio,Ciuchini:2017mik}. \\

In summary, our model is complementary and differs from previous works because it takes into account kinetic mixing and tries to address the LHCb anomaly while successfully hosting a fermionic dark matter particle (see \cite{Sierra:2015fma,Patra:2016shz,Kawamura:2017ecz,Cline:2017aed,Cline:2017lvv} for attempts to address simultaneously both of these issues in different models).\\

Our works is structured as follows: In {\it Section II} we describe the model, in {\it Section III} we derive the relevant interactions for our phenomenology, in {\it Section IV} we address the existing bounds on a $Z^\prime$ gauge boson arising from gauging the $L_\mu-L_\tau$ symmetry, in {\it Section V} we present our numerical results concerning dark matter and the B-decay anomaly, before concluding.

%%%%%%%%%%%%%%%%%%%%%%%%%%%%%%%%%%%%%%%%%%%%%%%%%%%%%%%%%%%%%%%%%%%%%%%%%%%%%%%%%%
\section{The Model}
\label{sec_model}
%%%%%%%%%%%%%%%%%%%%%%%%%%%%%%%%%%%%%%%%%%%%%%%%%%%%%%%%%%%%%%%%%%%%%%%%%%%%%%%%%%

In this paper we investigate a kinetically mixed $L_\mu - L_\tau$ symmetry~\cite{Altmannshofer:2014cfa,Altmannshofer:2016jzy}. In addition to the Standard Model (SM) terms, the model includes interactions of the new $\hat{Z}'$ gauge boson and a kinetic mixing term
\begin{equation}
\mathcal{L}_{\text{kin mix}} = -\frac{\sin(\epsilon)}{2} \hat{B}_{\mu \nu} \hat{Z}'^{\mu \nu},
\end{equation}
where the hats on the gauge fields denote that they are in the mixed basis and the kinetic mixing parameter $\sin(\epsilon)$ appears. The coupling of the new $U(1)'$ gauge boson to the SM is determined by the covariant derivative $D_\alpha = \partial_\alpha - i \tilde{g} q_{L_\mu - L_\tau} \hat{Z}'_\alpha$, with coupling strength $\tilde{g}$ and charge $q_{L_\mu - L_\tau}$ of the particle under the $U(1)_{L_\mu - L_\tau}$ gauge group, leading to
\begin{equation}
\mathcal{L}_{\ell Z'} = \tilde{g} \left( \bar{\ell}_2 \gamma^\alpha \ell_2 - \bar{\ell}_3 \gamma^\alpha \ell_3 + \bar{\mu}_R \gamma^\alpha \mu_R - \bar{\tau}_R \gamma^\alpha \tau_R \right) \hat{Z}'_\alpha,
\end{equation}
with $\ell_{2/3}$ denoting the left-handed lepton doublets of the second and third generation. This can be rewritten in the useful form
\begin{equation}
\label{eq_lepton_Zprime_interactions}
\mathcal{L}_{\ell Z'} = q_\ell \tilde{g} \left( \bar{\mu} \gamma^\alpha \mu - \bar{\tau} \gamma^\alpha \tau + \bar{\nu}_\mu \gamma^\alpha P_L \nu_\mu - \bar{\nu}_\tau \gamma^\alpha P_L \nu_\tau \right) \hat{Z}'_\alpha,
\end{equation}
with the projection operator $P_L \coloneqq \frac{1}{2}(1-\gamma^5)$ and we included an additional factor $q_\ell$ accompanying the $\tilde{g}$ in case the symmetry of the new $U(1)'$ is a multiple of $L_\mu - L_\tau$.\\

Dark Matter is introduced, in this setup, by considering a new particle charged under the new $U(1)^{'}$ symmetry, which also guarantees its stability, while being a SM singlet. One simple possibility is to consider a vector-like Dirac fermion $\chi$. Its coupling to the $\hat{Z}'$ is described by the following Lagrangian:
\begin{equation}
\mathcal{L}_{\text{dark}} = q_\chi \tilde{g} \bar{\chi} \gamma^\mu \chi \hat{Z}'_\mu,
\end{equation}
with charge $q_\chi$ of the DM under the new gauge group. While it would be intriguing to consider a dynamical generation of the DM mass $m_\chi$ from the spontaneous breaking of the $U(1)^{'}$ symmetry, we will postpone it to future study and just regard $m_\chi$ as a free parameter here.\\

Transforming to the standard kinetic terms and diagonalizing the mass matrix~\cite{Babu:1997st,Chun:2010ve}, we find for the relation of the gauge fields in the mixed basis to the physical fields (using the shorthand notation for trigonometric functions)
\begin{align}
\label{eq_mixed_to_physical_basis}
\begin{aligned}
\hat{W}_\mu^{+/-} &= W_\mu^{+/-} \\
\hat{A}_\mu &= A_\mu - \hat{c}_W t_\epsilon s_\alpha Z_\mu - \hat{c}_W t_\epsilon c_\alpha Z'_\mu \\
\hat{Z}_\mu &= \left( c_\alpha + \hat{s}_W t_\epsilon s_\alpha \right) Z_\mu + \left( \hat{s}_W t_\epsilon c_\alpha - s_\alpha \right) Z'_\mu \\
\hat{Z}'_\mu &= \frac{s_\alpha}{c_\epsilon} Z_\mu + \frac{c_\alpha}{c_\epsilon} Z'_\mu.
\end{aligned}
\end{align}
Furthermore, the diagonalization yields for the masses
\begin{align}
\label{eq_physical_masses}
\begin{aligned}
m_{Z'/Z}^2 &= \frac{1}{2} \hat{m}_Z^2 + \frac{\hat{m}_Z^2 \hat{s}_W^2 s_\epsilon^2 + \hat{m}_{Z'}^2}{2 c_\epsilon^2} \\
&\pm \sqrt{\left( \frac{1}{2} \hat{m}_Z^2 + \frac{\hat{m}_Z^2 \hat{s}_W^2 s_\epsilon^2 + \hat{m}_{Z'}^2}{2 c_\epsilon^2} \right)^2 - \frac{1}{c_\epsilon^2} \hat{m}_Z^2 \hat{m}_{Z'}^2}
\end{aligned}
\end{align}
and the relations of the mixing \mbox{angle $\alpha$} and the masses
\begin{equation}
\label{eq_mixing_angle}
c_\alpha^2 = \frac{\hat{m}_Z^2 - m_{Z'}^2}{m_Z^2 - m_{Z'}^2}
\qquad
\text{and}
\qquad
c_\alpha s_\alpha = \frac{\hat{m}_Z^2 \hat{s}_W t_\epsilon}{m_Z^2 - m_{Z'}^2}.
\end{equation}

From these equations, we can derive the useful identities
\begin{equation}
t_{2 \alpha} = -\frac{\hat{m}_Z^2 \hat{s}_W s_{2 \epsilon}}{\hat{m}_{Z'}^2 - \hat{m}_Z^2 (c_\epsilon^2 - s_\epsilon^2 \hat{s}_W^2)},
\end{equation}
for the mixing angle, and
\begin{equation}
\label{eq_physical_m_Z}
m_Z^2 = \hat{m}_Z^2 (1+\hat{s}_W t_\alpha t_\epsilon)
\end{equation}
and
\begin{equation}
m_{Z'}^2 = \frac{\hat{m}_{Z'}^2}{c_\epsilon^2 (1+\hat{s}_W t_\alpha t_\epsilon)},
\end{equation}
for the physical masses. Furthermore, we can define $r \coloneqq \frac{m_{Z'}}{m_Z}$, in our case $r>1$, and use Eqs.~\eqref{eq_mixing_angle} and~\eqref{eq_physical_m_Z} to get
\begin{equation}
t_\alpha = \frac{-r^2+1+\sqrt{(r^2-1)^2-4 \hat{s}_W^2 t_\epsilon^2 r^2}}{2 \hat{s}_W t_\epsilon r^2},
\end{equation}
where the sign in front of the square root is determined by the fact that we want $t_\alpha \rightarrow 0$ for $\epsilon \rightarrow 0$. For a realistic set of parameters this can be approximated to an accuracy of better than $1$ \textperthousand~by
\begin{equation}
\label{eq_mixng_angle_approximation}
t_\alpha \approx -\frac{\hat{s}_W t_\epsilon}{r^2 - 1}.
\end{equation}

%%%%%%%%%%%%%%%%%%%%%%%%%%%%%%%%%%%%%%%%%%%%%%%%%%%%%%%%%%%%%%%%%%%%%%%%%%%%%%%%%%
\section{Interactions, Cross-Sections and Decay Width}
\label{sec_interactions}
%%%%%%%%%%%%%%%%%%%%%%%%%%%%%%%%%%%%%%%%%%%%%%%%%%%%%%%%%%%%%%%%%%%%%%%%%%%%%%%%%%

Using the relations in Eq.~\eqref{eq_mixed_to_physical_basis}, we can determine the interactions of the SM states and the DM in the physical basis. In this basis the DM features interactions both with the $Z$ and the $Z^{'}$ boson given by:
\begin{equation}
\mathcal{L}_{\text{dark}} = g_{\chi Z} \bar{\chi} \gamma^\mu \chi Z_\mu + g_{\chi Z'} \bar{\chi} \gamma^\mu \chi Z'_\mu,
\end{equation}
where we defined
\begin{align}
\begin{aligned}
g_{\chi Z} &\coloneqq \frac{s_\alpha}{c_\epsilon} g_\chi, \\
g_{\chi Z'} &\coloneqq \frac{c_\alpha}{c_\epsilon} g_\chi,
\end{aligned}
\end{align}
with $g_\chi \coloneqq q_\chi \tilde{g}$.

The interaction of the SM fermions with the physical photon are left invariant by the change of basis, hence they read:
\begin{equation}
\mathcal{L}_A = \hat{e} \bar{f} \gamma^\mu Q A_\mu f,
\end{equation}
where we can make the identification $\hat{e} = e$ (cf.~\cite{Babu:1997st}) and $Q$ is the electric charge of the fermion $f$. The same holds true for the $W$, which as well has unaltered charged current interactions. Concerning the neutral current interactions we have:
\begin{align}
\begin{aligned}
\mathcal{L}_{NC,Z} &= \bar{f} \gamma^\mu \left( g_{{fZ}_L} P_L + g_{{fZ}_R} P_R \right) f Z_\mu, \\
\mathcal{L}_{NC,Z'} &= \bar{f} \gamma^\mu \left( g_{{fZ'}_L} P_L + g_{{fZ'}_R} P_R \right) f Z'_\mu.
\end{aligned}
\end{align}
While for the case of quarks and first generation leptons these couplings depend only on SM parameters as well as the mixing parameters $\epsilon$ and $\alpha$:
\begin{align}
\begin{aligned}
g_{{fZ}_L} &\coloneqq \frac{e}{\hat{s}_W \hat{c}_W} c_\alpha [T_3 (1 + \hat{s}_W t_\epsilon t_\alpha) - Q (\hat{s}_W^2 + \hat{s}_W t_\epsilon t_\alpha)], \\
g_{{fZ}_R} &\coloneqq -\frac{e}{\hat{s}_W \hat{c}_W} c_\alpha Q (\hat{s}_W^2 + \hat{s}_W t_\epsilon t_\alpha), \\
g_{{fZ'}_L} &\coloneqq \frac{e}{\hat{s}_W \hat{c}_W} c_\alpha [T_3 (\hat{s}_W t_\epsilon - t_\alpha) + Q (\hat{s}_W^2 t_\alpha - \hat{s}_W t_\epsilon)], \\
g_{{fZ'}_R} &\coloneqq \frac{e}{\hat{s}_W \hat{c}_W} c_\alpha Q (\hat{s}_W^2 t_\alpha - \hat{s}_W t_\epsilon),
\end{aligned}
\end{align}
there are additional contribution from Eq.~\eqref{eq_lepton_Zprime_interactions} in the case of second and third generation leptons, so that:
\begin{align}
\begin{aligned}
g_{{\mu/\tau Z}_L} &\coloneqq -\frac{e}{2 \hat{s}_W \hat{c}_W} c_\alpha [(1 + \hat{s}_W t_\epsilon t_\alpha) - 2 (\hat{s}_W^2 + \hat{s}_W t_\epsilon t_\alpha)] \\
&\quad \pm g_\ell \frac{s_\alpha}{c_\epsilon}, \\
g_{{\mu/\tau Z}_R} &\coloneqq \frac{e}{\hat{s}_W \hat{c}_W} c_\alpha (\hat{s}_W^2 + \hat{s}_W t_\epsilon t_\alpha) \pm g_\ell \frac{s_\alpha}{c_\epsilon}, \\
g_{{\nu_{\mu/\tau} Z}_L} &\coloneqq \frac{e}{2 \hat{s}_W \hat{c}_W} c_\alpha (1 + \hat{s}_W t_\epsilon t_\alpha) \pm g_\ell \frac{s_\alpha}{c_\epsilon}, \\
g_{{\nu_{\mu/\tau} Z}_R} &\coloneqq 0,
\end{aligned}
\end{align}
for the $Z$, where the plus sign refers to the muon and the minus sign to the tau, while for the $Z'$ boson we have:
\begin{align}
\begin{aligned}
g_{{\mu/\tau Z'}_L} &\coloneqq -\frac{e}{2 \hat{s}_W \hat{c}_W} c_\alpha [(\hat{s}_W t_\epsilon - t_\alpha) + 2 (\hat{s}_W^2 t_\alpha - \hat{s}_W t_\epsilon)] \\
&\quad \pm g_\ell \frac{c_\alpha}{c_\epsilon}, \\
g_{{\mu/\tau Z'}_R} &\coloneqq -\frac{e}{\hat{s}_W \hat{c}_W} c_\alpha (\hat{s}_W^2 t_\alpha - \hat{s}_W t_\epsilon) \pm g_\ell \frac{c_\alpha}{c_\epsilon}, \\
g_{{\nu_{\mu/\tau} Z'}_L} &\coloneqq \frac{e}{2 \hat{s}_W \hat{c}_W} c_\alpha (\hat{s}_W t_\epsilon - t_\alpha) \pm g_\ell \frac{c_\alpha}{c_\epsilon}, \\
g_{{\nu_{\mu/\tau} Z'}_R} &\coloneqq 0,
\end{aligned}
\end{align}
where we have introduced $g_\ell \coloneqq q_\ell \tilde{g}$.

Gauge self interactions can be straightforwardly determined by realizing that all terms that appear for the photon $A_\mu$ are also there for the $Z_\mu$ with an additional prefactor of $\frac{\hat{c}_W}{\hat{s}_W}$. This, together with
\begin{equation}
\hat{A}_\mu + \frac{\hat{c}_W}{\hat{s}_W} \hat{Z}_\mu = A_\mu + \frac{\hat{c}_W}{\hat{s}_W} c_\alpha Z_\mu - \frac{\hat{c}_W}{\hat{s}_W} s_\alpha Z'_\mu,
\end{equation}
shows that the photon gauge interactions do not change, while the $Z'$ interactions get weighted by a factor of $c_\alpha$ and also appear for the $Z'$ weighted by a factor of $-s_\alpha$. This leads for the relevant three point interactions to
\begin{equation}
\mathcal{L}_{VWW} = g_{ZWW} [[Z W W]] + g_{Z'WW} [[Z' W W]],
\end{equation}
where we abbreviated $[[V W W]] \equiv -i[(\partial_\mu W_\nu^+ - \partial_\nu W_\mu^+) W^{- \mu} V^\nu - (\partial_\mu W_\nu^- - \partial_\nu W_\mu^-) W^{+ \mu} V^\nu + \frac{1}{2} (\partial_\mu V_\nu - \partial_\nu V_\mu)(W^{+\mu} W^{-\nu} - W^{-\mu} W^{+\nu})]$, and defined
\begin{align}
\begin{aligned}
g_{ZWW} &\coloneqq \frac{e}{\hat{t}_W} c_\alpha, \\
g_{Z'WW} &\coloneqq -\frac{e}{\hat{t}_W} s_\alpha.
\end{aligned}
\end{align}

The last relevant kind of interactions are the ones with the Higgs, represented by
\begin{equation}
\mathcal{L}_{hVV} = \frac{1}{2} g_{ZZh} h Z_\mu Z^\mu + \frac{1}{2} g_{Z'Z'h} h Z'_\mu Z'^\mu + g_{ZZ'h} h Z_\mu Z'^\mu,
\end{equation}
where we defined
\begin{align}
\begin{aligned}
g_{ZZh} &\coloneqq \frac{e}{\hat{s}_W \hat{c}_W^2} c_\alpha^2 m_W (1 + \hat{s}_W t_\epsilon t_\alpha)^2, \\
g_{Z'Z'h} &\coloneqq \frac{e}{\hat{s}_W \hat{c}_W^2} c_\alpha^2 m_W (\hat{s}_W t_\epsilon - t_\alpha)^2, \\
g_{ZZ'h} &\coloneqq \frac{e}{\hat{s}_W \hat{c}_W^2} c_\alpha^2 m_W (\hat{s}_W t_\epsilon + \hat{s}_W^2 t_\epsilon^2 t_\alpha - t_\alpha - \hat{s}_W t_\epsilon t_\alpha^2).
\end{aligned}
\end{align}
This concludes the list of relevant interactions and it is worth mentioning that we neglected four point interactions for the Higgs and gauge self interactions, since they are not relevant at tree level for the processes we take into account.

\begin{figure}[h]
\includegraphics[width=\columnwidth]{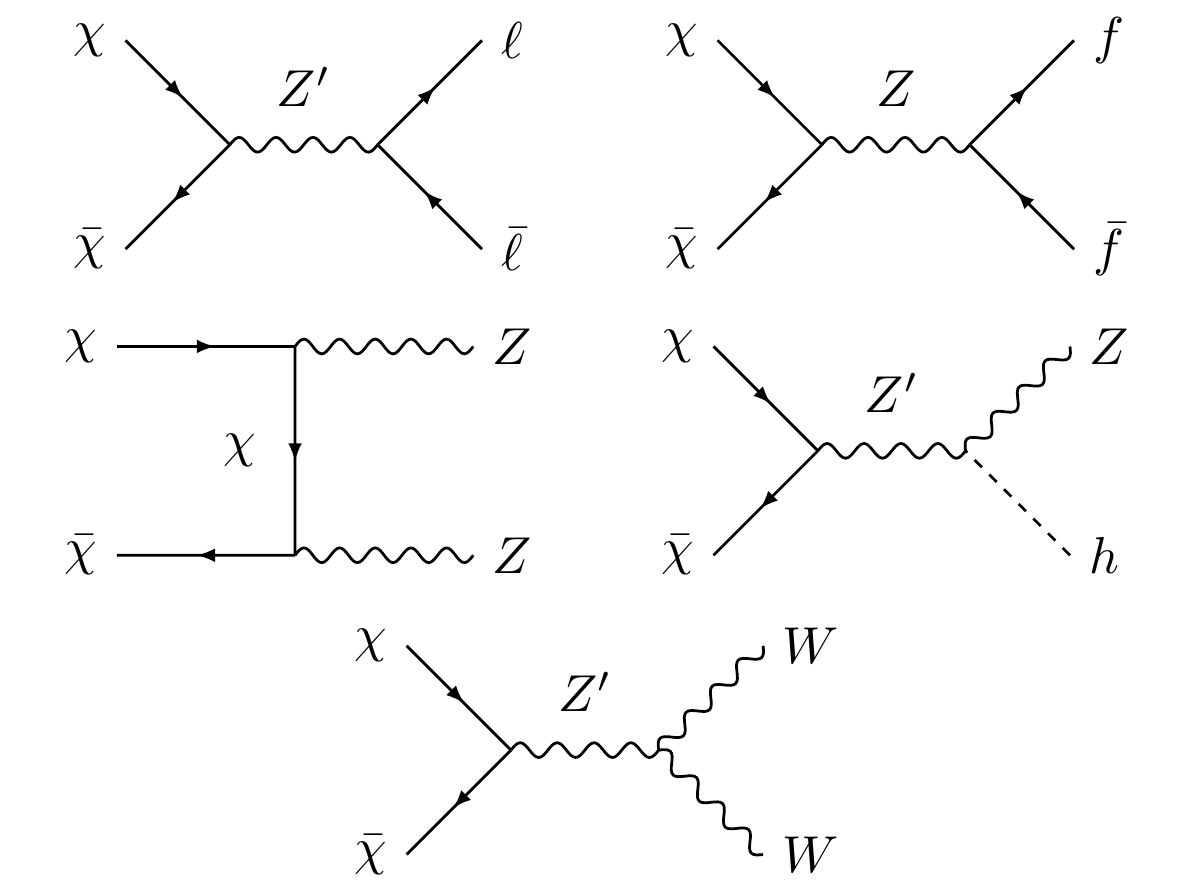}
\caption{Feynman diagrams that contribute to the relic density calculation. The first diagram accounts for s-channel annihilation via the $Z^\prime$ into leptons charged under $L_\mu-L_\tau$; the second arises via the $Z-Z^\prime$ kinetic mixing allowing dark matter to annihilate into all SM fermions; the third represents the t-channel annihilations into $Z\,Z$, $Z\,Z^\prime$ or $Z^\prime Z^\prime$ pairs; the forth and fifth also resulted from the $Z-Z^\prime$ mixing and deal with annihilations into $W\, W$ and $Z\,h$.}
\label{fenmanndiag}
\end{figure}

With the interactions at hand we can now compute the dark matter annihilation cross section and the decay width of the new $Z^\prime$ boson. The Feynman diagrams for the possible dark matter annihilation channels are depicted in Fig.~\ref{fenmanndiag}.\\

For the dark matter annihilation we adopted the conventional velocity expansion, $\langle \sigma v \rangle \simeq a+b v^2$, retaining only the leading order term, thus getting relatively simple analytical expressions:

\begin{widetext}

\begin{align}
\label{eq:sigmas}
\begin{aligned}
\langle \sigma v \rangle_{\chi \chi \rightarrow f f} &= \frac{N_c}{2 \pi} m_\chi^2 \sqrt{1 - \frac{m_f^2}{m_\chi^2}} \left[ (g_{f_L}^2 + g_{f_R}^2) \left( 1 - \frac{m_f^2}{4 m_\chi^2} \right) + \frac{3}{2} g_{f_L} g_{f_R} \frac{m_f^2}{m_\chi^2} \right], \\
\langle \sigma v \rangle_{\chi \chi \rightarrow W^+ W^-} &= \frac{1}{\pi} m_\chi^2 g_W^2 \left( 1 - \frac{m_W^2}{m_\chi^2} \right)^{\frac{3}{2}} \left[ \frac{m_\chi^4}{m_W^4} + 5 \frac{m_\chi^2}{m_W^2} + \frac{3}{4} \right], \\
\langle \sigma v \rangle_{\chi \chi \rightarrow Z Z} &= \frac{g_{\chi Z}^4}{16 \pi m_\chi^2} \left( 1 - \frac{m_Z^2}{m_\chi^2} \right)^{\frac{3}{2}} \left( 1 - \frac{m_Z^2}{2 m_\chi^2} \right)^{-2}, \\
\langle \sigma v \rangle_{\chi \chi \rightarrow Z Z'} &= \frac{g_{\chi Z}^2 g_{\chi Z'}^2}{8 \pi m_\chi^2} \left( \frac{(m_Z^2-m_{Z'}^2)^2}{16 m_\chi^4} + 1 - \frac{m_Z^2 + m_{Z'}^2}{2 m_\chi^2} \right)^{\frac{3}{2}} \left( 1 - \frac{m_Z^2 + m_{Z'}^2}{4 m_\chi^2} \right)^{-2}, \\
\langle \sigma v \rangle_{\chi \chi \rightarrow Z h} &= \frac{1}{8 \pi} g_h^2 \sqrt{1 - \frac{m_h^2 + m_Z^2}{2 m_\chi^2} + \left( \frac{m_h^2 - m_Z^2}{4 m_\chi^2} \right)^2} \left[ 1 + \frac{1}{2} \frac{m_\chi^2}{m_Z^2} \left( 1 - \frac{m_h^2 - m_Z^2}{2 m_\chi^2} + \left( \frac{m_h^2 - m_Z^2}{4 m_\chi^2} \right) \right) \right],
\end{aligned}
\end{align}

\end{widetext}

where the decay $\chi \chi \rightarrow Z' Z'$ is identical to the one to $Z Z$ only with every $Z$ replaced by $Z'$ and for the decay $\chi \chi \rightarrow Z' h$ in comparison to the one to $Z h$ one has to replace $Z$ by $Z'$ and vice versa (also in $g_h$). The factor $N_c$ accounts for particle multiplicity, i.e. the different colors of the quarks. To be able to display the expressions in a more compact way we defined
\begin{align}
\begin{aligned}
g_{f_L} &\coloneqq \frac{g_{\chi Z} g_{{fZ}_L}}{4m_\chi^2 - m_Z^2} + \frac{g_{\chi Z'} g_{{f Z'}_L}}{4m_\chi^2 - m_Z'^2}, \\
g_{f_R} &\coloneqq \frac{g_{\chi Z} g_{{fZ}_R}}{4m_\chi^2 - m_Z^2} + \frac{g_{\chi Z'} g_{{f Z'}_R}}{4m_\chi^2 - m_Z'^2}, \\
g_W &\coloneqq \frac{g_{\chi Z} g_{ZWW}}{4m_\chi^2 - m_Z^2} + \frac{g_{\chi Z'} g_{Z'WW}}{4m_\chi^2 - m_Z'^2}, \\
g_h &\coloneqq \frac{g_{\chi Z} g_{ZZh}}{4m_\chi^2 - m_Z^2} + \frac{g_{\chi Z'} g_{Z'Zh}}{4m_\chi^2 - m_Z'^2}.
\end{aligned}
\end{align}

Similar calculations for the $Z'$ decay width lead to

\begin{widetext}

\begin{align}
\begin{aligned}
\Gamma_{Z' \rightarrow \chi \chi} &= \frac{g_{\chi Z'}^2}{6 \pi} \sqrt{\frac{m_{Z'}^2}{4} - m_\chi^2} \left( 1+ 2 \frac{m_\chi^2}{m_{Z'}^2} \right), \\
\Gamma_{Z' \rightarrow f f} &= \frac{N_c}{12 \pi} \sqrt{\frac{m_{Z'}^2}{4} - m_f^2} \left[ \left(g_{{f Z'}_L}^2 + g_{{f Z'}_R}^2\right) \left( 1 - \frac{m_f^2}{m_{Z'}^2} \right) + 6 g_{{f Z'}_L} g_{{f Z'}_R} \frac{m_f^2}{m_{Z'}^2} \right], \\
\Gamma_{Z' \rightarrow W^+ W^-} &= \frac{g_{ZWW}^2}{192 \pi} m_{Z'} \left(\frac{m_{Z'}}{m_W}\right)^4 \left( 1 - 4 \frac{m_W^2}{m_{Z'}^2} \right)^{\frac{3}{2}} \left[ 1 + 20 \frac{m_W^2}{m_{Z'}^2} + 12 \frac{m_W^4}{m_{Z'}^4} \right], \\
\Gamma_{Z' \rightarrow h Z} &= \frac{g_{Z Z' h}^2}{48 \pi m_{Z'}^3} \sqrt{(m_h^2 + m_Z^2 - m_{Z'}^2)^2 - 4 m_h^2 m_Z^2} \left[ 2 + \frac{1}{4} \left( \frac{m_{Z'}}{m_Z} + \frac{m_Z}{m_{Z'}} - \frac{m_h^2}{m_Z m_{Z'}} \right)^2 \right].
\end{aligned}
\end{align}

\end{widetext}

Now that we have derived the relevant interactions for our model, we present the bounds on the parameter space as summarized in the next section.

%%%%%%%%%%%%%%%%%%%%%%%%%%%%%%%%%%%%%%%%%%%%%%%%%%%%%%%%%%%%%%%%%%%%%%%%%%%%%%%%%%
\section{Bounds on the Model Parameters}
\label{sec_bounds}
%%%%%%%%%%%%%%%%%%%%%%%%%%%%%%%%%%%%%%%%%%%%%%%%%%%%%%%%%%%%%%%%%%%%%%%%%%%%%%%%%%
In this section we will, besides the LHC bounds, address two constraints that refer to electroweak precision and neutrino trident production. They restrict three key parameters in our study namely, the kinetic mixing, the $Z^\prime$ mass and the $L_\mu-L_\tau$ gauge coupling. We begin with electroweak precision which is imprinted on the $\rho$ parameter.

\subsection{Electroweak Precision}

To relate the $\rho$ parameter to the model, however, we will need to make a small detour and have a closer look at the weak mixing (or Weinberg) angle first. Since it is the angle that determines the rotation from the $\hat{W}_\mu^3$ and $\hat{B}_\mu$ to the $\hat{Z}_\mu$ and $\hat{A}_\mu$ fields, it is affected by the diagonalization of the $Z$-$Z'$-sector. Therefore, there are two weak mixing angles. One is the usual one from the SM defined by
\begin{equation}
\label{eq_def_c_W}
\hat{c}_W = \frac{\hat{m}_W}{\hat{m}_Z},
\end{equation}
and is denoted with a hat since it belongs to the mixed basis. The other one is the physical weak mixing angle (the one measured in a lab), which is conventionally defined as~\cite{Babu:1997st}
\begin{equation}
\label{eq_physical_weak_mixing_angle}
s_W^2 c_W^2 = \frac{\pi \alpha(m_Z)}{\sqrt{2} G_F m_Z^2},
\end{equation}
where $\alpha(m_Z)$ is the weak coupling constant at the $Z$-pole and $G_F$ is the Fermi constant. The Eq.~\eqref{eq_physical_weak_mixing_angle} holds also in the mixed basis with $s_W \rightarrow \hat{s}_W$, $c_W \rightarrow \hat{c}_W$ and $m_Z \rightarrow \hat{m}_Z$, so we get~\cite{Babu:1997st}
\begin{equation}
\label{eq_weak_mixing_angle_basis_relation}
s_W^2 c_W^2 m_Z^2 = \hat{s}_W^2 \hat{c}_W^2 \hat{m}_Z^2.
\end{equation}
Using Eq.~\eqref{eq_physical_m_Z} we find
\begin{equation}
s_W^2 c_W^2 = \frac{\hat{s}_W^2 \hat{c}_W^2}{1 + \hat{s}_W t_\alpha t_\epsilon},
\end{equation}
which leads to a relation of $\hat{s}_W$ and $s_W$ by plugging in the approximation of Eq.~\eqref{eq_mixng_angle_approximation} and $c_W^2 = 1 - s_W^2$. Since for $\epsilon \rightarrow 0$ we want $\hat{s}_W = s_W$, we get

\begin{widetext}

\begin{equation}
\label{eq_s_W_relation}
\hat{s}_W^2 \approx \frac{1}{2} \left[ 1 + \frac{s_W^2 t_\epsilon^2}{r^2-1} - \frac{s_W^4 t_\epsilon^2}{r^2-1} - \sqrt{\left(1 + \frac{s_W^2 t_\epsilon^2}{r^2-1} - \frac{s_W^4 t_\epsilon^2}{r^2-1}\right)^2 - 4 (s_W^2 - s_W^4)} \right].
\end{equation}

\end{widetext}

This relation enables us to determine $\hat{s}_W$ from the model parameters $s_\epsilon$ and $m_{Z'}$ and the experimentally determined quantities $s_W$ and $m_Z$. To get a better idea of how $\hat{s}_W$ qualitatively behaves we can expand this expression in $\epsilon$ around $0$ to get
\begin{equation}
\hat{s}_W^2 \approx s_W^2 \left( 1 - \frac{s_W^2 c_W^2 \epsilon^2}{(c_W^2-s_W^2)(r^2-1)} \right) + \mathcal{O}(\epsilon^4),
\end{equation}
however we will continue to work with the more exact expression. \\

\begin{figure}[h]
\centering
\includegraphics[width=\columnwidth]{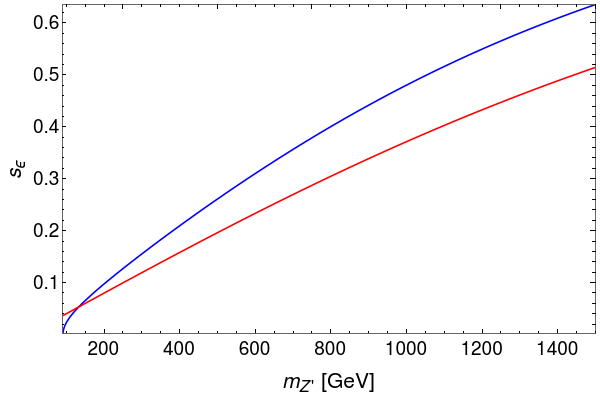}
\caption{Maximal possible kinetic mixing parameter depending on the mass of the new gauge boson $m_{Z'}$. In blue the limit from the $2 \sigma$ upper bound of the $\rho$ parameter is shown, while in red the limit from EWPT~\cite{Kumar:2006gm} is depicted.}
\label{fig_kinetic_mixing_limits}
\end{figure}

With this knowledge at hand we can derive a limit on the mixing angle from the measurement of the $\rho$-parameter. By definition we have
\begin{equation}
\rho = \frac{m_W^2}{m_Z^2 c_W^2},
\end{equation}
which can with $m_W = \hat{m}_W$ and Eq.~\eqref{eq_weak_mixing_angle_basis_relation} be rewritten to
\begin{equation}
\rho = \frac{\hat{m}_W^2}{\hat{m}_Z^2 \hat{c}_W^2} \frac{s_W^2}{\hat{s}_W^2} = \frac{s_W^2}{\hat{s}_W^2}
\end{equation}
because the first fraction is by definition of $\hat{c}_W$ identical to one (cf. Eq.~\eqref{eq_def_c_W}). Together with Eq.~\eqref{eq_s_W_relation} for $\hat{s}_W^2$, this enables us to determine a limit on $s_\epsilon$ depending on the value of $m_{Z'}$ - by using the PDG$16$ values~\cite{Patrignani:2016xqp}. In addition to the limit from the $\rho$ parameter, we can also use the limit form electroweak precision tests (EWPT)~\cite{Kumar:2006gm}
\begin{equation}
\left(\frac{t_\epsilon}{0.1}\right)^2 \left(\frac{250\,\text{GeV}}{m_{Z'}}\right)^2 \lesssim 1.
\end{equation}
Both of these limits are shown in Fig.~\ref{fig_kinetic_mixing_limits} and one can see that except for a small interval below $m_{Z'}\sim 130\,\text{GeV}$ the bound from EWPT is stronger.

This already shrinks the parameter space of the model by limiting the possible values for the kinetic mixing parameter.  We will move on to neutrino trident production, which will allow us to constraint the $Z^\prime$ mass and the gauge coupling. 

\subsection{Neutrino Trident Production}

\begin{figure}[h]
\centering
\includegraphics[width=0.6\columnwidth]{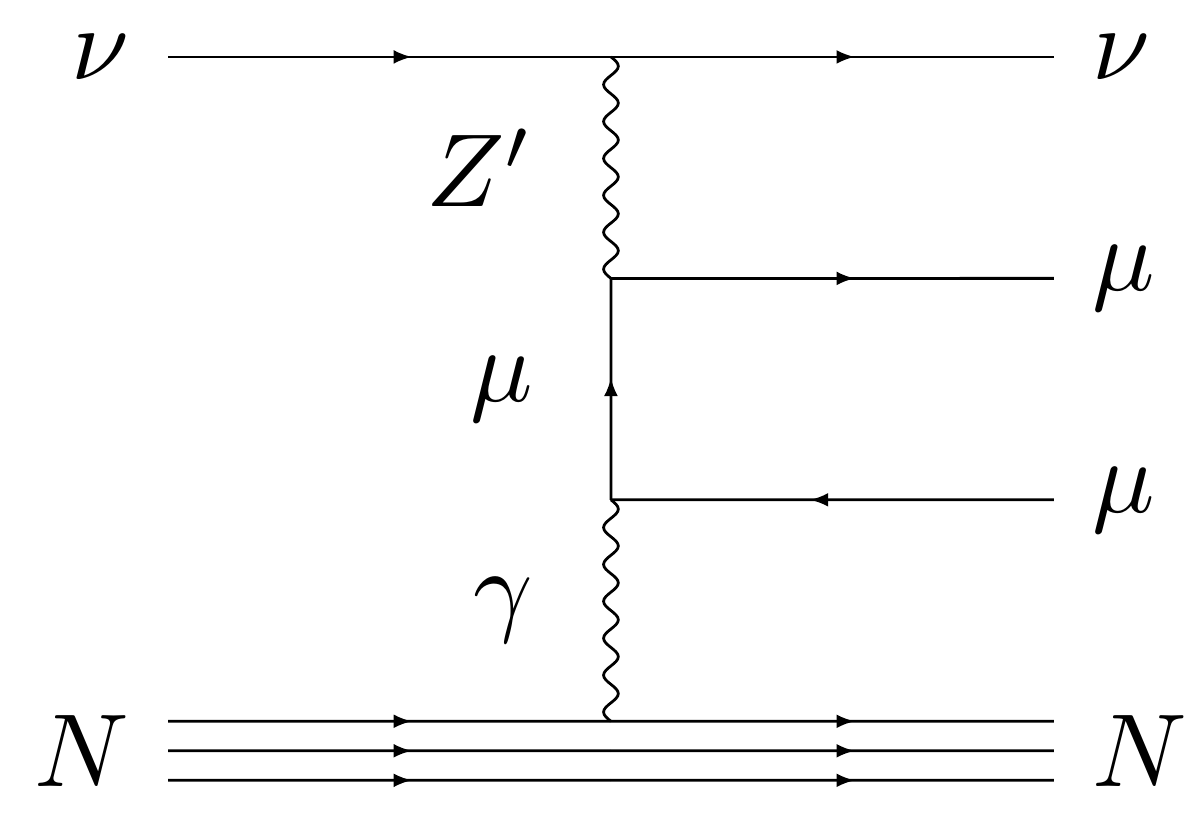}
\caption{Feynman diagram for the leading $Z^\prime$ contribution to neutrino trident production.}
\label{figtrident}
\end{figure}

Neutrino trident production refers to the scattering of a neutrino off a heavy nucleus where a pair of muons is produced in the final state \cite{Altmannshofer:2014cfa,Ge:2017poy} (the leading $Z^\prime$ contribution is shown in Fig.~\ref{figtrident}). If this process can, in addition to $Z$ interactions, also happen via $Z^\prime$ interactions to the neutrinos and muons, the expected rate changes compared to the SM. For this reason, neutrino trident production is a perfect laboratory to probe the $L_\mu-L_\tau$ symmetry. The process has been measured and due to the reasonable agreement with SM predictions one can use this observation to constrain muonic forces such as the one present in the $L_\mu-L_\tau$ model. This has been carried out in~\cite{Altmannshofer:2014pba}, where a lower mass bound of $m_{Z^\prime} > 540\ \mbox{GeV} g_l$ was found. The corresponding bound is shown as a hatched region in the figures. Only one coupling appears in this lower mass bound because the $Z^\prime$ couples with equal strength, $g_l$, to the leptons. Furthermore, the analysis in~\cite{Altmannshofer:2014pba} also shows that it is not possible for a heavy $Z^\prime$ ($ M_{Z^\prime} \gtrsim 400\,\text{MeV}$) to explain the muon $g-2$ discrepancy within an $L_\mu - L_\tau$ model without kinetic mixing. This should remain true also after including kinetic mixing, since none of the relevant couplings is significantly suppressed or enhanced because the mixing parameter is restricted to be small (cf. Fig.~\ref{fig_kinetic_mixing_limits}).

\subsection{B-Anomaly and LHC bounds}

Over the last few years anomalies have been observed in rare B meson decays hinting at new physics~\cite{Blake:2016olu}. There are uncertainties surrounding the SM prediction for these decays, but global fits seem to favor new physics effects in the $b\rightarrow s \mu\mu$ transitions \cite{Altmannshofer:2014rta,Descotes-Genon:2015uva}. Describing this using effective field theory, the data favor $C_9^{NP}\simeq -1.07$, where the coefficient rescales the effective Lagrangian
\begin{equation}
L_{NP}= \frac{- 4 G_F \alpha_{em}}{\sqrt{2} 4\pi}C_9^{NP} \left[V_{tb}V^{\ast}_{ts} (\bar{s} \gamma_\alpha P_L b)(\bar{\mu}\gamma^\alpha \mu)\right].
\end{equation}

If we add this effective term to our model in the attempt to explain the LHCb anomaly, we automatically also alter the $B_s-\bar{B}_s$ mixing amplitude as predicted by the SM and which has been measured. Allowing the deviation from the SM prediction to lie within 15\% \cite{Charles:2015gya,Bazavov:2016nty}, the upper bound $m_{Z^{\prime}} < 4.9\,\mbox{TeV} g_l (-1.07/C_p^{NP})$ is found \cite{Altmannshofer:2016jzy}.\\

Having in mind the bound we presented stemming from neutrino-trident production, we conclude that the favored region to explain the B-anomaly is $540\,\mbox{GeV} < m_{Z^{\prime}}/\,g_l < 4.9\,\mbox{TeV}$. This band is labeled as {\it B anomaly} in the figures.  \\

Due to the kinetic mixing, the $Z'$ features couplings with quarks and can hence be resonantly produced in proton-proton collisions. Among the possible decay final states the strongest constraints are provided by the searches for dilepton final states. We have translated the limits from most recent searches~\cite{Khachatryan:2015dcf,Aaboud:2017buh} into excluded regions in the two-dimensional plane $(m_{Z'},m_\chi)$ for some assignations of the parameters $s_\epsilon,g_l,g_\chi$. The limits are strongest for $m_{Z'} < 2 m_\chi$, while otherwise they are weakened by the presence of the ``invisible'' decay branching fraction for the $Z'$ into DM pairs. \\

These are not the only relevant bounds for the model. The other limits are related to dark matter searches, which we will address below.

\begin{figure}[h]
\centering
\includegraphics[width=\columnwidth]{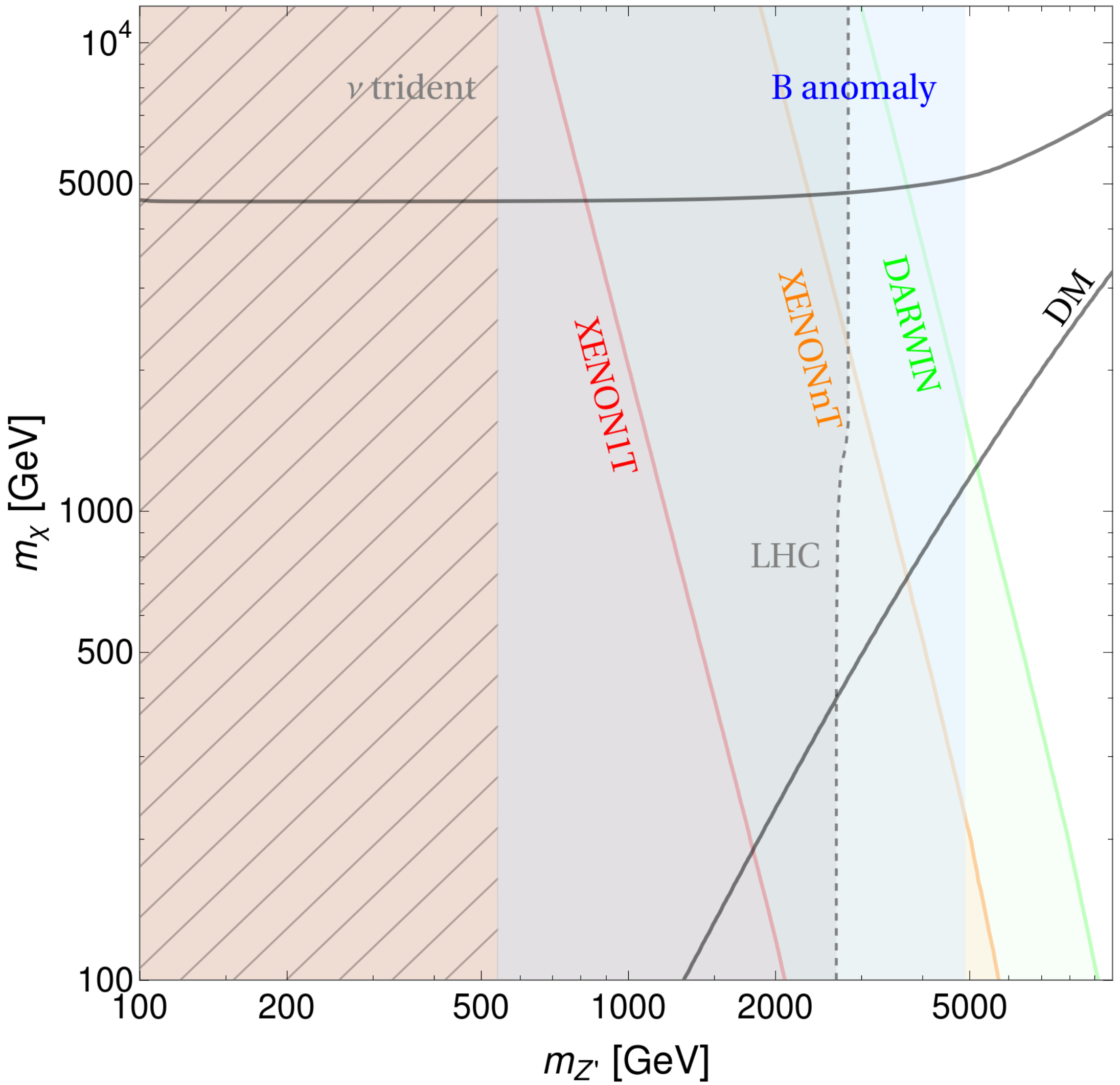}
\caption{Exclusion plot of the $Z'$ mass mixing model for a kinetic mixing parameter of $s_\epsilon = 0.1$ and couplings of $g_\ell = 1$ and $g_\chi = 1$. The black curve marks the parameter points with the correct relic DM density. The red area is excluded by XENON1T and the orange and green area show the reach of XENONnT and DARWIN. The hatched region is excluded by neutrino trident production, the gray region up to the dashed line is ruled out by dilepton searches at the LHC and the blue region is favored by measurements of B meson anomalies.}
\label{fig_exclusion_plot_se01_gl1_gchi1}
\end{figure}

\begin{figure}[h]
\centering
\includegraphics[width=\columnwidth]{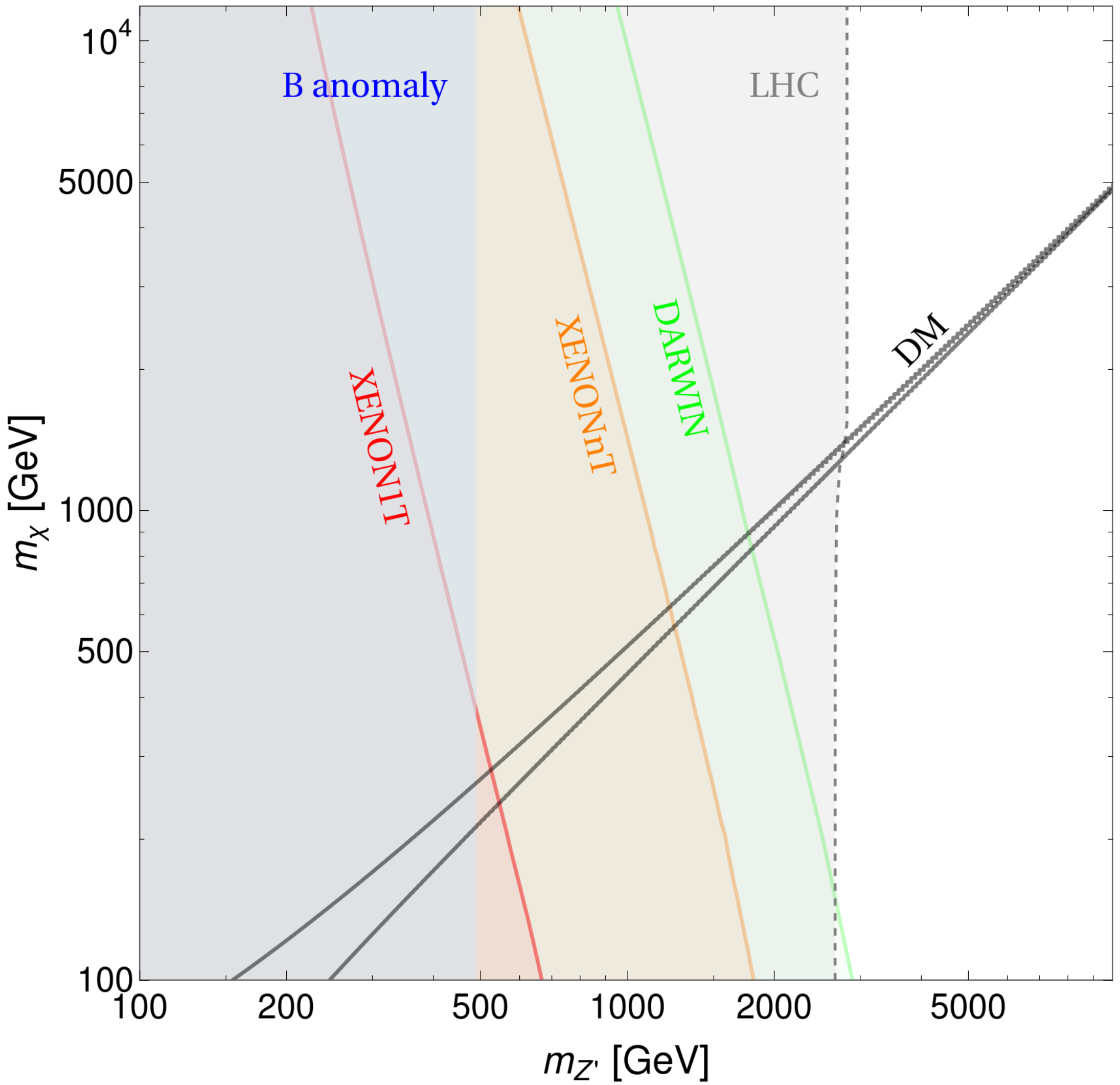}
\caption{Exclusion plot of the $Z'$ mass mixing model for a kinetic mixing parameter of $s_\epsilon = 0.1$ and couplings of $g_\ell = 0.1$ and $g_\chi = 0.1$. The black curve marks the parameter points with the correct relic DM density. The red area is excluded by XENON1T and the orange and green area show the reach of XENONnT and DARWIN. The hatched region is excluded by neutrino trident production, the gray region up to the dashed line is ruled out by dilepton searches at the LHC and the blue region is favored by measurements of B meson anomalies.}
\label{fig_exclusion_plot_se01_gl01_gchi01}
\end{figure}

\section{Dark Matter Phenomenology}

In this section we address the dark matter phenomenology of our model in the presence of a fermionic dark matter particle in the context of the WIMP miracle \cite{Arcadi:2017kky} whose interactions are dictated by the gauge symmetry $L_\mu -L_\tau$ and the kinetic mixing parameter. \\

The first key observable to be discussed is the dark matter relic density which can be computed using the thermally averaged annihilations cross sections given in Sec.~\ref{sec_interactions} as a first approximation and away from the $Z^\prime$ resonance. While the velocity expansion is very useful for illustration, it is not valid in some potentially relevant regimes~\cite{Griest:1990kh}, like the $m_\chi \sim m_{Z'}/2$ resonance region. For this reason, our results are based on a more rigorous numerical treatment performed by implementing the model in the micrOMEGAs package \cite{Belanger:2013oya}. In this way we have derived the relic density curve shown in the figures with a {\it solid black curve} by finding the parameter space that yields $\Omega_{\text{DM}} h^2 = 0.12$ as measured by the \textsc{Planck} collaboration~\cite{Ade:2015xua}~\footnote{We have assumed thermal freeze-out throughout and ignored any non-standard cosmology effects \cite{Dutta:2016htz,Dutra:2018gmv}}.\\

After checking that our fermionic DM candidate can achieve the experimental determined abundance, we need to asses whether it also obeys the restrictive limits from direct detection experiments \cite{Angle:2011th,Akerib:2015rjg,Aguilar-Arevalo:2016zop,Aguilar-Arevalo:2016ndq,Tan:2016zwf,Agnese:2017njq,Amole:2017dex,Aprile:2017iyp,Aprile:2017ngb,Aprile:2017yea,Cui:2017nnn,Akerib:2017kat,Agnes:2018oej,Agnes:2018ves}. Regardless of the null results, these experiments have played a crucial role in our understanding of dark matter properties since they have ruled out several dark matter models by limiting the maximal dark matter nucleon scattering cross-section. \\

In our model, the dark fermion scatters off nuclei at tree-level via the $Z-Z^\prime$ mixing. The Feynman diagram for this process is the t-channel version of the second diagram in Fig.~\ref{fenmanndiag} with either a $Z$ or a $Z^\prime$ as mediator. Therefore, our model is subject to bounds stemming from direct detection experiments. The relevant limits come from spin-independent (SI) interactions and are usually expressed as upper limits on the scattering cross section of the DM on protons, as function of its mass. This upper limit should be compared with the theoretical prediction: 
\begin{equation}
\label{eq_SI_cross-section}
\sigma_{\text{SI}} (\chi p \rightarrow \chi p) = \frac{1}{\pi} \mu_{\chi, p}^2 f_p^2 \frac{\left[ Z  + (A-Z) \frac{f_n}{f_p} \right]^2}{A^2}.
\end{equation}
where $\mu_{\chi, p} = \frac{m_\chi m_p}{m_\chi + m_p}$. $f_p$ and $f_n$ represent the effective couplings of the DM with, respectively, protons and neutrons, and are given by:
\begin{align}
& f_p=\frac{g_{\chi Z} \left(2 (g_{u Z_L}+g_{u Z_R})+(g_{d Z_L}+g_{d Z_R})\right)}{4 m_Z^2}\nonumber\\
& +\frac{g_{\chi Z'} \left(2 (g_{u Z'_L}+g_{u Z'_R})+(g_{d Z'_L}+g_{d Z'_R})\right)}{4 m_Z^{'\,2}}\nonumber\\
& f_n=\frac{g_{\chi Z} \left((g_{u Z_L}+g_{u Z_R})+2 (g_{d Z_L}+g_{d Z_R})\right)}{4 m_Z^2}\nonumber\\
& +\frac{g_{\chi Z'} \left((g_{u Z'_L}+g_{u Z'_R})+2(g_{d Z'_L}+g_{d Z'_R})\right)}{4 m_Z^{'\,2}}.
\end{align}
Note that, while the theoretical scattering cross-section of the DM on protons would just read $\frac{1}{\pi} \mu_{\chi, N}^2 f_p^2$, we need to introduce, for a direct comparison with the experimental limit, the factor $\left[ Z  + (A-Z) \frac{f_n}{f_p} \right]^2 / A^2$. Indeed, once converting the limits on the DM scattering rate on nuclei into limits on the scattering cross-section of the DM on protons, it is conventionally assumed that the DM interacts with the same strength with protons and nucleons. This is assumption is not valid for the model under consideration; consequently we had to introduce a normalization factor for the cross-section in order to achieve a proper comparison with experimental constraints (see e.g.~\cite{Feng:2013fyw} for more details). \\

The strongest limits on the spin-independent dark matter-nucleon scattering cross section are given by XENON1T~\cite{Aprile:2017iyp}, while a slightly weaker constrained is provided by PANDA-X \cite{Cui:2017nnn} (all these experiments are based on Xenon detectors, hence $Z=54$, while we have adopted $A=131$). Given the similarity of the limits we will refer to them as a unique curve labeled XENON1T. We will also present the projected limits from XENONnT \cite{Aprile:2015uzo} and DARWIN \cite{Aalbers:2016jon} which use the same material and readout techniques. Notice that these exclusion limits have been determined by assuming that the experimental value of the DM local density is always adopted in computing the DM scattering rate on the various detectors. This implies the implicit assumption that the DM candidate under consideration is the only DM component of the Universe and features the correct relic density $\Omega_{\text{DM}} h^2 = 0.12$ irrespective of the parameters of the theory. In other words we are assuming that, outside the contours corresponding to the correct relic density according to the WIMP paradigm, the correct DM abundance is accommodated, for example, by some non thermal production mechanism or, possibly, modified cosmological history of the Universe. \\

An important observable related to the dark matter relic density is the dark matter annihilation cross section. Indirect detection experiments such as Fermi-LAT, AMS-02, and H.E.S.S. and CTA provide stringent bounds on the dark matter annihilation cross section \cite{Hooper:2012sr,DeAngelis:2017gra,Acharya:2017ttl,Li:2018kgy}. However, these bounds are generally more relevant for dark matter masses below $\sim 100$~GeV in agreement with \cite{Bringmann:2014lpa,Ackermann:2015zua,Giesen:2015ufa,Profumo:2017obk,Balazs:2017hxh}. Therefore, for the parameter space this work is focused on, indirect detection is left aside. \\

In the following, we present exclusion plots for the model for a variety of parameter choices. They are supposed to give an impression of where in the parameter space the model is still viable. The constraints are due to the fact that we want to get the correct dark matter relic density while still being compatible with dark matter direct detection experiments, neutrino trident production and dilepton searches at the LHC. In all exclusion plots the black curve shows the points in the parameter space that reproduce the correct DM relic density - with the area in between having a too small one and outside it is too large - and the colored regions show the excluded area due to the direct detection experiment limits - with the red area being excluded by XENON1T, the orange area giving the reach of XENONnT and the green area the projected sensitivity of DARWIN. In blue we overlay the region in which the {\it B anomaly} could be potentially addressed using the effective description aforementioned. The hatched region refers to neutrino trident production limits (\textit{$\nu$ trident}) and the dashed line to \textit{LHC} dilepton searches.\\

\begin{figure}[h]
\centering
\includegraphics[width=\columnwidth]{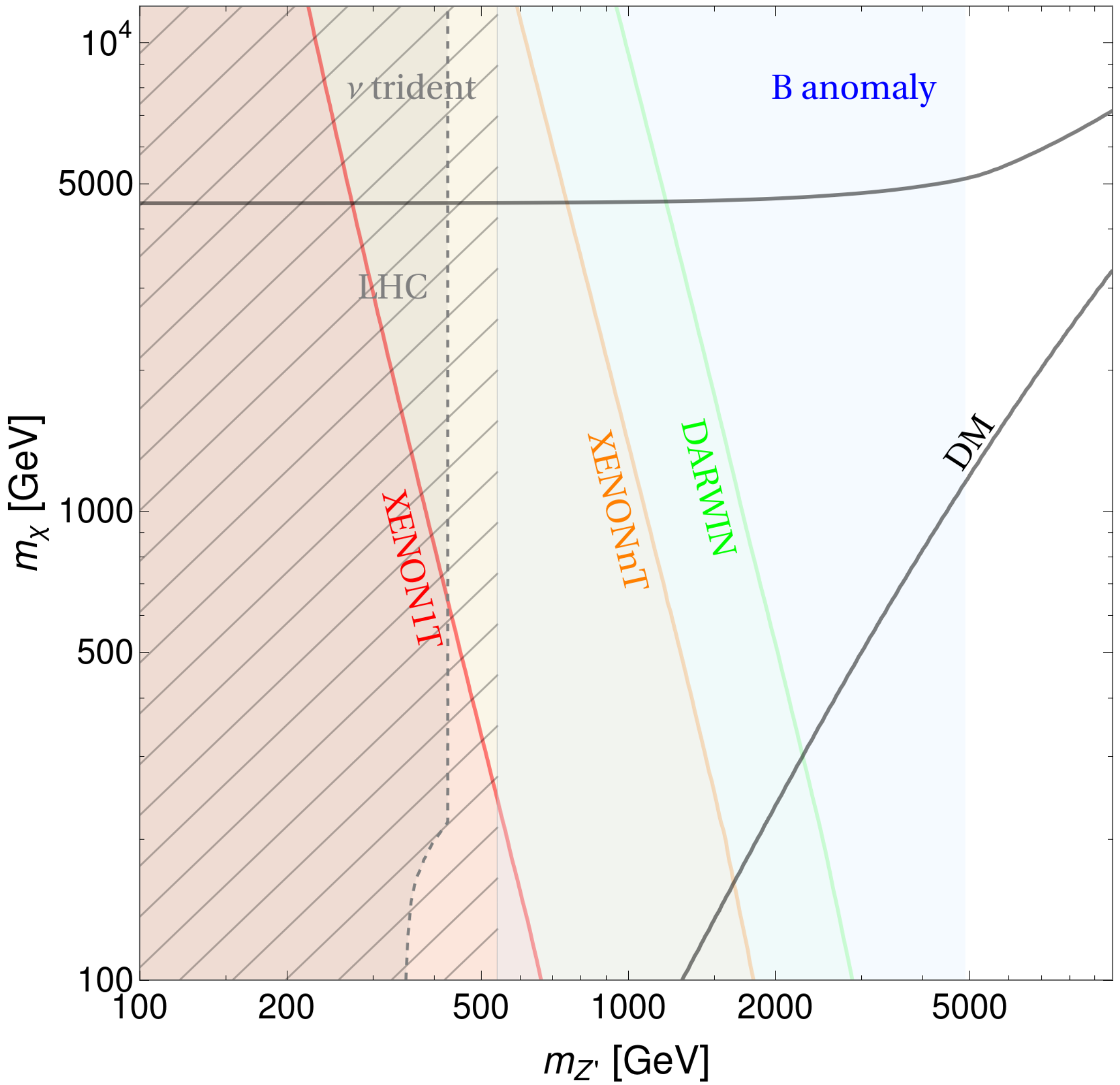}
\caption{Exclusion plot of the $Z'$ mass mixing model for a kinetic mixing parameter of $s_\epsilon = 0.01$ and couplings of $g_\ell = 1$ and $g_\chi = 1$. The black curve marks the parameter points with the correct relic DM density. The red area is excluded by XENON1T and the orange and green area show the reach of XENONnT and DARWIN. The hatched region is excluded by neutrino trident production, the gray region up to the dashed line is ruled out by dilepton searches at the LHC and the blue region is favored by measurements of B meson anomalies.}
\label{fig_exclusion_plot_se001_gl1_gchi1}
\end{figure}

There are several aspects that are apparent in the plots shown in Figs.~\ref{fig_exclusion_plot_se01_gl1_gchi1},~\ref{fig_exclusion_plot_se01_gl01_gchi01} and~\ref{fig_exclusion_plot_se001_gl1_gchi1}. The most effective bounds are by far the ones from XENON1T and LHC searches of dilepton resonances. The latter, in particular, provide for $s_\epsilon=0.1$ the strongest exclusion limits. This can be understood from the fact that even if the production cross section is moderately suppressed by the kinetic mixing parameter, the direct coupling of the $Z'$ with the muons enhances its decay branching fractions into dileptons. The production cross section of the $Z'$ becomes nevertheless rapidly suppressed as $s_\epsilon$ decreases. For $s_\epsilon=0.01$ the exclusion bound becomes indeed sub-dominant with respect to the one from neutrino trident production. \\

Nevertheless, regions of the parameter space which can reproduce the correct relic density, compatible with exclusion bounds, are present. Notice that in Figs.~\ref{fig_exclusion_plot_se01_gl1_gchi1} and~\ref{fig_exclusion_plot_se001_gl1_gchi1} we have used $g_{\chi}=1$ while in Fig.~\ref{fig_exclusion_plot_se01_gl01_gchi01} we adopted $g_\chi=0.1$. Having that in mind, when $m_{\chi} > m_{Z^\prime}$ and $g_{\chi}$ is sufficiently large, the t-channel annihilation into $Z^\prime Z^\prime$ drives the annihilation rate and for this reason the relic density curves in Figs.~\ref{fig_exclusion_plot_se01_gl1_gchi1} and~\ref{fig_exclusion_plot_se001_gl1_gchi1} exhibit a similar behavior. This fact allows one to have $Z^\prime$ masses below $1$~TeV in agreement with direct detection data while still reproducing the correct relic density.\\

Furthermore, XENONnT and DARWIN will rule out significant parts of the parameter space (XENONnT leads to a roughly three times larger limit on the $Z'$ mass for the same DM mass and DARWIN excludes even more). A small $g_\ell$ makes it harder to reproduce the correct DM relic density (one needs to be ``more resonant''), while not giving an advantage in terms of direct detection bounds. Also, a small $g_\chi$ helps to evade direct detection bounds, but makes it harder to get the correct DM relic density (a small kinetic mixing parameter is better in that regard). Moreover, one can notice that in Figs.~\ref{fig_exclusion_plot_se01_gl1_gchi1} and~\ref{fig_exclusion_plot_se001_gl1_gchi1} one can simultaneously accommodate dark matter and the {\it B anomaly} for TeV dark matter. When the dark matter particle is sufficiently heavy the t-channel annihilation into $Z^\prime Z^\prime$ opens up changing the shape of the relic density curve while not suffering from strong limits from direct detection. For this reason our setup favors heavy dark matter, different to what occurred in \cite{Altmannshofer:2016jzy}. Anyways, it is clear that in case we continue to observe no signal of dark matter scattering, the next generation of experiments will restrict the model to live in a corner of the parameter space.

%%%%%%%%%%%%%%%%%%%%%%%%%%%%%%%%%%%%%%%%%%%%%%%%%%%%%%%%%%%%%%%%%%%%%%%%%%%%%%%%%%
\section{Conclusion}
\label{sec_conclusion}
%%%%%%%%%%%%%%%%%%%%%%%%%%%%%%%%%%%%%%%%%%%%%%%%%%%%%%%%%%%%%%%%%%%%%%%%%%%%%%%%%%

In this work we have investigated the Dark Matter phenomenology in a $L_{\mu}-L_{\tau}$ extension of the SM, allowing in addition for kinetic mixing between the new $Z^{\prime}$ and the SM $Z$ boson. We have compared the requirement of correct relic density, according to the WIMP paradigm, with constraints/prospects from DM direct searches and collider searches for the new $Z^\prime$ boson. Furthermore, we have examined whether the considered setup can account for the {\it B anomaly}, within the allowed parameter space from DM phenomenology. \\

We revisited limits on the $L_\mu-L_\tau$ model such as those from neutrino trident production and concluded that the most competitive constraint comes from DM direct detection, which can probe even relatively small values of the kinetic mixing parameter. Larger values of the kinetic mixing parameter are restricted most strongly by LHC dilepton searches. For order 1 values of the gauge coupling $g_l$ associated to the new gauge symmetry, it is possible to account for the {\it B anomaly} compatibly with the correct DM relic density and still evading limits from DM Direct Detection. Next generation experiments, like XENONnT and DARWIN will noticeably reduce the region where dark matter and the {\it B anomaly} are simultaneously accounted for, in case of null results.

%%%%%%%%%%%%%%%%%%%%%%%%%%%%%%%%%%%%%%%%%%%%%%%%%%%%%%%%%%%%%%%%%%%%%%%%%%%%%%%%%%
\acknowledgments
%%%%%%%%%%%%%%%%%%%%%%%%%%%%%%%%%%%%%%%%%%%%%%%%%%%%%%%%%%%%%%%%%%%%%%%%%%%%%%%%%%

We would like to thank Miguel D. Campos, Carlos Pires and Alexander J. Helmboldt for helpful and interesting discussions. FSQ acknowledges financial support from MEC, UFRN and ICTP-SAIFR FAPESP grant 2016/01343-7.

\bibliography{literature}

\begin{thebibliography}{105}%
\makeatletter
\providecommand \@ifxundefined [1]{%
 \@ifx{#1\undefined}
}%
\providecommand \@ifnum [1]{%
 \ifnum #1\expandafter \@firstoftwo
 \else \expandafter \@secondoftwo
 \fi
}%
\providecommand \@ifx [1]{%
 \ifx #1\expandafter \@firstoftwo
 \else \expandafter \@secondoftwo
 \fi
}%
\providecommand \natexlab [1]{#1}%
\providecommand \enquote  [1]{``#1''}%
\providecommand \bibnamefont  [1]{#1}%
\providecommand \bibfnamefont [1]{#1}%
\providecommand \citenamefont [1]{#1}%
\providecommand \href@noop [0]{\@secondoftwo}%
\providecommand \href [0]{\begingroup \@sanitize@url \@href}%
\providecommand \@href[1]{\@@startlink{#1}\@@href}%
\providecommand \@@href[1]{\endgroup#1\@@endlink}%
\providecommand \@sanitize@url [0]{\catcode `\\12\catcode `\$12\catcode
  `\&12\catcode `\#12\catcode `\^12\catcode `\_12\catcode `\%12\relax}%
\providecommand \@@startlink[1]{}%
\providecommand \@@endlink[0]{}%
\providecommand \url  [0]{\begingroup\@sanitize@url \@url }%
\providecommand \@url [1]{\endgroup\@href {#1}{\urlprefix }}%
\providecommand \urlprefix  [0]{URL }%
\providecommand \Eprint [0]{\href }%
\providecommand \doibase [0]{http://dx.doi.org/}%
\providecommand \selectlanguage [0]{\@gobble}%
\providecommand \bibinfo  [0]{\@secondoftwo}%
\providecommand \bibfield  [0]{\@secondoftwo}%
\providecommand \translation [1]{[#1]}%
\providecommand \BibitemOpen [0]{}%
\providecommand \bibitemStop [0]{}%
\providecommand \bibitemNoStop [0]{.\EOS\space}%
\providecommand \EOS [0]{\spacefactor3000\relax}%
\providecommand \BibitemShut  [1]{\csname bibitem#1\endcsname}%
\let\auto@bib@innerbib\@empty
%</preamble>
\bibitem [{\citenamefont {Ade}\ \emph {et~al.}(2016)\citenamefont {Ade} \emph
  {et~al.}}]{Ade:2015xua}%
  \BibitemOpen
  \bibfield  {author} {\bibinfo {author} {\bibfnamefont {P.~A.~R.}\
  \bibnamefont {Ade}} \emph {et~al.} (\bibinfo {collaboration} {Planck}),\
  }\href {\doibase 10.1051/0004-6361/201525830} {\bibfield  {journal} {\bibinfo
   {journal} {Astron. Astrophys.}\ }\textbf {\bibinfo {volume} {594}},\
  \bibinfo {pages} {A13} (\bibinfo {year} {2016})},\ \Eprint
  {http://arxiv.org/abs/1502.01589} {arXiv:1502.01589 [astro-ph.CO]}
  \BibitemShut {NoStop}%
%%CITATION = ARXIV:1502.01589;%%
\bibitem [{\citenamefont {Queiroz}\ \emph {et~al.}(2017)\citenamefont
  {Queiroz}, \citenamefont {Rodejohann},\ and\ \citenamefont
  {Yaguna}}]{Queiroz:2016sxf}%
  \BibitemOpen
  \bibfield  {author} {\bibinfo {author} {\bibfnamefont {F.~S.}\ \bibnamefont
  {Queiroz}}, \bibinfo {author} {\bibfnamefont {W.}~\bibnamefont {Rodejohann}},
  \ and\ \bibinfo {author} {\bibfnamefont {C.~E.}\ \bibnamefont {Yaguna}},\
  }\href {\doibase 10.1103/PhysRevD.95.095010} {\bibfield  {journal} {\bibinfo
  {journal} {Phys. Rev.}\ }\textbf {\bibinfo {volume} {D95}},\ \bibinfo {pages}
  {095010} (\bibinfo {year} {2017})},\ \Eprint
  {http://arxiv.org/abs/1610.06581} {arXiv:1610.06581 [hep-ph]} \BibitemShut
  {NoStop}%
%%CITATION = ARXIV:1610.06581;%%
\bibitem [{\citenamefont {Catena}\ \emph {et~al.}(2017)\citenamefont {Catena},
  \citenamefont {Conrad},\ and\ \citenamefont {Krauss}}]{Catena:2017xqq}%
  \BibitemOpen
  \bibfield  {author} {\bibinfo {author} {\bibfnamefont {R.}~\bibnamefont
  {Catena}}, \bibinfo {author} {\bibfnamefont {J.}~\bibnamefont {Conrad}}, \
  and\ \bibinfo {author} {\bibfnamefont {M.~B.}\ \bibnamefont {Krauss}},\
  }\href@noop {} {\  (\bibinfo {year} {2017})},\ \Eprint
  {http://arxiv.org/abs/1712.07969} {arXiv:1712.07969 [hep-ph]} \BibitemShut
  {NoStop}%
%%CITATION = ARXIV:1712.07969;%%
\bibitem [{\citenamefont {Kavanagh}\ \emph {et~al.}(2017)\citenamefont
  {Kavanagh}, \citenamefont {Queiroz}, \citenamefont {Rodejohann},\ and\
  \citenamefont {Yaguna}}]{Kavanagh:2017hcl}%
  \BibitemOpen
  \bibfield  {author} {\bibinfo {author} {\bibfnamefont {B.~J.}\ \bibnamefont
  {Kavanagh}}, \bibinfo {author} {\bibfnamefont {F.~S.}\ \bibnamefont
  {Queiroz}}, \bibinfo {author} {\bibfnamefont {W.}~\bibnamefont {Rodejohann}},
  \ and\ \bibinfo {author} {\bibfnamefont {C.~E.}\ \bibnamefont {Yaguna}},\
  }\href {\doibase 10.1007/JHEP10(2017)059} {\bibfield  {journal} {\bibinfo
  {journal} {JHEP}\ }\textbf {\bibinfo {volume} {10}},\ \bibinfo {pages} {059}
  (\bibinfo {year} {2017})},\ \Eprint {http://arxiv.org/abs/1706.07819}
  {arXiv:1706.07819 [hep-ph]} \BibitemShut {NoStop}%
%%CITATION = ARXIV:1706.07819;%%
\bibitem [{\citenamefont {Ackermann}\ \emph {et~al.}(2015)\citenamefont
  {Ackermann} \emph {et~al.}}]{Ackermann:2015zua}%
  \BibitemOpen
  \bibfield  {author} {\bibinfo {author} {\bibfnamefont {M.}~\bibnamefont
  {Ackermann}} \emph {et~al.} (\bibinfo {collaboration} {Fermi-LAT}),\ }\href
  {\doibase 10.1103/PhysRevLett.115.231301} {\bibfield  {journal} {\bibinfo
  {journal} {Phys. Rev. Lett.}\ }\textbf {\bibinfo {volume} {115}},\ \bibinfo
  {pages} {231301} (\bibinfo {year} {2015})},\ \Eprint
  {http://arxiv.org/abs/1503.02641} {arXiv:1503.02641 [astro-ph.HE]}
  \BibitemShut {NoStop}%
%%CITATION = ARXIV:1503.02641;%%
\bibitem [{\citenamefont {Wood}\ \emph {et~al.}(2016)\citenamefont {Wood},
  \citenamefont {Anderson}, \citenamefont {Drlica-Wagner}, \citenamefont
  {Cohen-Tanugi},\ and\ \citenamefont {Conrad}}]{Wood:2015ofa}%
  \BibitemOpen
  \bibfield  {author} {\bibinfo {author} {\bibfnamefont {M.}~\bibnamefont
  {Wood}}, \bibinfo {author} {\bibfnamefont {B.}~\bibnamefont {Anderson}},
  \bibinfo {author} {\bibfnamefont {A.}~\bibnamefont {Drlica-Wagner}}, \bibinfo
  {author} {\bibfnamefont {J.}~\bibnamefont {Cohen-Tanugi}}, \ and\ \bibinfo
  {author} {\bibfnamefont {J.}~\bibnamefont {Conrad}} (\bibinfo {collaboration}
  {Fermi-LAT}),\ }\bibfield  {booktitle} {\emph {\bibinfo {booktitle}
  {{Proceedings, 34th International Cosmic Ray Conference (ICRC 2015): The
  Hague, The Netherlands, July 30-August 6, 2015}}},\ }\href@noop {} {\bibfield
   {journal} {\bibinfo  {journal} {PoS}\ }\textbf {\bibinfo {volume}
  {ICRC2015}},\ \bibinfo {pages} {1226} (\bibinfo {year} {2016})},\ \Eprint
  {http://arxiv.org/abs/1507.03530} {arXiv:1507.03530 [astro-ph.HE]}
  \BibitemShut {NoStop}%
%%CITATION = ARXIV:1507.03530;%%
\bibitem [{\citenamefont {Abdallah}\ \emph {et~al.}(2016)\citenamefont
  {Abdallah} \emph {et~al.}}]{Abdallah:2016ygi}%
  \BibitemOpen
  \bibfield  {author} {\bibinfo {author} {\bibfnamefont {H.}~\bibnamefont
  {Abdallah}} \emph {et~al.} (\bibinfo {collaboration} {H.E.S.S.}),\ }\href
  {\doibase 10.1103/PhysRevLett.117.111301} {\bibfield  {journal} {\bibinfo
  {journal} {Phys. Rev. Lett.}\ }\textbf {\bibinfo {volume} {117}},\ \bibinfo
  {pages} {111301} (\bibinfo {year} {2016})},\ \Eprint
  {http://arxiv.org/abs/1607.08142} {arXiv:1607.08142 [astro-ph.HE]}
  \BibitemShut {NoStop}%
%%CITATION = ARXIV:1607.08142;%%
\bibitem [{\citenamefont {Ahnen}\ \emph {et~al.}(2016)\citenamefont {Ahnen}
  \emph {et~al.}}]{Ahnen:2016qkx}%
  \BibitemOpen
  \bibfield  {author} {\bibinfo {author} {\bibfnamefont {M.~L.}\ \bibnamefont
  {Ahnen}} \emph {et~al.} (\bibinfo {collaboration} {Fermi-LAT, MAGIC}),\
  }\href {\doibase 10.1088/1475-7516/2016/02/039} {\bibfield  {journal}
  {\bibinfo  {journal} {JCAP}\ }\textbf {\bibinfo {volume} {1602}},\ \bibinfo
  {pages} {039} (\bibinfo {year} {2016})},\ \Eprint
  {http://arxiv.org/abs/1601.06590} {arXiv:1601.06590 [astro-ph.HE]}
  \BibitemShut {NoStop}%
%%CITATION = ARXIV:1601.06590;%%
\bibitem [{\citenamefont {Archambault}\ \emph {et~al.}(2017)\citenamefont
  {Archambault} \emph {et~al.}}]{Archambault:2017wyh}%
  \BibitemOpen
  \bibfield  {author} {\bibinfo {author} {\bibfnamefont {S.}~\bibnamefont
  {Archambault}} \emph {et~al.} (\bibinfo {collaboration} {VERITAS}),\ }\href
  {\doibase 10.1103/PhysRevD.95.082001} {\bibfield  {journal} {\bibinfo
  {journal} {Phys. Rev.}\ }\textbf {\bibinfo {volume} {D95}},\ \bibinfo {pages}
  {082001} (\bibinfo {year} {2017})},\ \Eprint
  {http://arxiv.org/abs/1703.04937} {arXiv:1703.04937 [astro-ph.HE]}
  \BibitemShut {NoStop}%
%%CITATION = ARXIV:1703.04937;%%
\bibitem [{\citenamefont {Foot}(1991)}]{Foot:1990mn}%
  \BibitemOpen
  \bibfield  {author} {\bibinfo {author} {\bibfnamefont {R.}~\bibnamefont
  {Foot}},\ }\href {\doibase 10.1142/S0217732391000543} {\bibfield  {journal}
  {\bibinfo  {journal} {Mod. Phys. Lett.}\ }\textbf {\bibinfo {volume} {A6}},\
  \bibinfo {pages} {527} (\bibinfo {year} {1991})}\BibitemShut {NoStop}%
%%CITATION = MPLAE,A6,527;%%
\bibitem [{\citenamefont {Foot}\ \emph {et~al.}(1994)\citenamefont {Foot},
  \citenamefont {He}, \citenamefont {Lew},\ and\ \citenamefont
  {Volkas}}]{Foot:1994vd}%
  \BibitemOpen
  \bibfield  {author} {\bibinfo {author} {\bibfnamefont {R.}~\bibnamefont
  {Foot}}, \bibinfo {author} {\bibfnamefont {X.~G.}\ \bibnamefont {He}},
  \bibinfo {author} {\bibfnamefont {H.}~\bibnamefont {Lew}}, \ and\ \bibinfo
  {author} {\bibfnamefont {R.~R.}\ \bibnamefont {Volkas}},\ }\href {\doibase
  10.1103/PhysRevD.50.4571} {\bibfield  {journal} {\bibinfo  {journal} {Phys.
  Rev.}\ }\textbf {\bibinfo {volume} {D50}},\ \bibinfo {pages} {4571} (\bibinfo
  {year} {1994})},\ \Eprint {http://arxiv.org/abs/hep-ph/9401250}
  {arXiv:hep-ph/9401250 [hep-ph]} \BibitemShut {NoStop}%
%%CITATION = HEP-PH/9401250;%%
\bibitem [{\citenamefont {Lindner}\ \emph {et~al.}(2018)\citenamefont
  {Lindner}, \citenamefont {Platscher},\ and\ \citenamefont
  {Queiroz}}]{Lindner:2016bgg}%
  \BibitemOpen
  \bibfield  {author} {\bibinfo {author} {\bibfnamefont {M.}~\bibnamefont
  {Lindner}}, \bibinfo {author} {\bibfnamefont {M.}~\bibnamefont {Platscher}},
  \ and\ \bibinfo {author} {\bibfnamefont {F.~S.}\ \bibnamefont {Queiroz}},\
  }\href {\doibase 10.1016/j.physrep.2017.12.001} {\bibfield  {journal}
  {\bibinfo  {journal} {Phys. Rep.}\ } (\bibinfo {year} {2018}),\
  10.1016/j.physrep.2017.12.001},\ \Eprint {http://arxiv.org/abs/1610.06587}
  {arXiv:1610.06587 [hep-ph]} \BibitemShut {NoStop}%
%%CITATION = ARXIV:1610.06587;%%
\bibitem [{\citenamefont {Heeck}\ and\ \citenamefont
  {Rodejohann}(2011)}]{Heeck:2011wj}%
  \BibitemOpen
  \bibfield  {author} {\bibinfo {author} {\bibfnamefont {J.}~\bibnamefont
  {Heeck}}\ and\ \bibinfo {author} {\bibfnamefont {W.}~\bibnamefont
  {Rodejohann}},\ }\href {\doibase 10.1103/PhysRevD.84.075007} {\bibfield
  {journal} {\bibinfo  {journal} {Phys. Rev.}\ }\textbf {\bibinfo {volume}
  {D84}},\ \bibinfo {pages} {075007} (\bibinfo {year} {2011})},\ \Eprint
  {http://arxiv.org/abs/1107.5238} {arXiv:1107.5238 [hep-ph]} \BibitemShut
  {NoStop}%
%%CITATION = ARXIV:1107.5238;%%
\bibitem [{\citenamefont {Freitas}\ \emph {et~al.}(2014)\citenamefont
  {Freitas}, \citenamefont {Lykken}, \citenamefont {Kell},\ and\ \citenamefont
  {Westhoff}}]{Freitas:2014pua}%
  \BibitemOpen
  \bibfield  {author} {\bibinfo {author} {\bibfnamefont {A.}~\bibnamefont
  {Freitas}}, \bibinfo {author} {\bibfnamefont {J.}~\bibnamefont {Lykken}},
  \bibinfo {author} {\bibfnamefont {S.}~\bibnamefont {Kell}}, \ and\ \bibinfo
  {author} {\bibfnamefont {S.}~\bibnamefont {Westhoff}},\ }\href {\doibase
  10.1007/JHEP09(2014)155, 10.1007/JHEP05(2014)145} {\bibfield  {journal}
  {\bibinfo  {journal} {JHEP}\ }\textbf {\bibinfo {volume} {05}},\ \bibinfo
  {pages} {145} (\bibinfo {year} {2014})},\ \bibinfo {note} {[Erratum:
  JHEP09,155(2014)]},\ \Eprint {http://arxiv.org/abs/1402.7065}
  {arXiv:1402.7065 [hep-ph]} \BibitemShut {NoStop}%
%%CITATION = ARXIV:1402.7065;%%
\bibitem [{\citenamefont {Altmannshofer}\ and\ \citenamefont
  {Yavin}(2015)}]{Altmannshofer:2015mqa}%
  \BibitemOpen
  \bibfield  {author} {\bibinfo {author} {\bibfnamefont {W.}~\bibnamefont
  {Altmannshofer}}\ and\ \bibinfo {author} {\bibfnamefont {I.}~\bibnamefont
  {Yavin}},\ }\href {\doibase 10.1103/PhysRevD.92.075022} {\bibfield  {journal}
  {\bibinfo  {journal} {Phys. Rev.}\ }\textbf {\bibinfo {volume} {D92}},\
  \bibinfo {pages} {075022} (\bibinfo {year} {2015})},\ \Eprint
  {http://arxiv.org/abs/1508.07009} {arXiv:1508.07009 [hep-ph]} \BibitemShut
  {NoStop}%
%%CITATION = ARXIV:1508.07009;%%
\bibitem [{\citenamefont {Omura}\ \emph {et~al.}(2016)\citenamefont {Omura},
  \citenamefont {Senaha},\ and\ \citenamefont {Tobe}}]{Omura:2015xcg}%
  \BibitemOpen
  \bibfield  {author} {\bibinfo {author} {\bibfnamefont {Y.}~\bibnamefont
  {Omura}}, \bibinfo {author} {\bibfnamefont {E.}~\bibnamefont {Senaha}}, \
  and\ \bibinfo {author} {\bibfnamefont {K.}~\bibnamefont {Tobe}},\ }\href
  {\doibase 10.1103/PhysRevD.94.055019} {\bibfield  {journal} {\bibinfo
  {journal} {Phys. Rev.}\ }\textbf {\bibinfo {volume} {D94}},\ \bibinfo {pages}
  {055019} (\bibinfo {year} {2016})},\ \Eprint
  {http://arxiv.org/abs/1511.08880} {arXiv:1511.08880 [hep-ph]} \BibitemShut
  {NoStop}%
%%CITATION = ARXIV:1511.08880;%%
\bibitem [{\citenamefont {Altmannshofer}\ \emph
  {et~al.}(2016{\natexlab{a}})\citenamefont {Altmannshofer}, \citenamefont
  {Carena},\ and\ \citenamefont {Crivellin}}]{Altmannshofer:2016oaq}%
  \BibitemOpen
  \bibfield  {author} {\bibinfo {author} {\bibfnamefont {W.}~\bibnamefont
  {Altmannshofer}}, \bibinfo {author} {\bibfnamefont {M.}~\bibnamefont
  {Carena}}, \ and\ \bibinfo {author} {\bibfnamefont {A.}~\bibnamefont
  {Crivellin}},\ }\href {\doibase 10.1103/PhysRevD.94.095026} {\bibfield
  {journal} {\bibinfo  {journal} {Phys. Rev.}\ }\textbf {\bibinfo {volume}
  {D94}},\ \bibinfo {pages} {095026} (\bibinfo {year} {2016}{\natexlab{a}})},\
  \Eprint {http://arxiv.org/abs/1604.08221} {arXiv:1604.08221 [hep-ph]}
  \BibitemShut {NoStop}%
%%CITATION = ARXIV:1604.08221;%%
\bibitem [{\citenamefont {Ibe}\ \emph {et~al.}(2017)\citenamefont {Ibe},
  \citenamefont {Nakano},\ and\ \citenamefont {Suzuki}}]{Ibe:2016dir}%
  \BibitemOpen
  \bibfield  {author} {\bibinfo {author} {\bibfnamefont {M.}~\bibnamefont
  {Ibe}}, \bibinfo {author} {\bibfnamefont {W.}~\bibnamefont {Nakano}}, \ and\
  \bibinfo {author} {\bibfnamefont {M.}~\bibnamefont {Suzuki}},\ }\href
  {\doibase 10.1103/PhysRevD.95.055022} {\bibfield  {journal} {\bibinfo
  {journal} {Phys. Rev.}\ }\textbf {\bibinfo {volume} {D95}},\ \bibinfo {pages}
  {055022} (\bibinfo {year} {2017})},\ \Eprint
  {http://arxiv.org/abs/1611.08460} {arXiv:1611.08460 [hep-ph]} \BibitemShut
  {NoStop}%
%%CITATION = ARXIV:1611.08460;%%
\bibitem [{\citenamefont {Baek}(2017)}]{Baek:2017sew}%
  \BibitemOpen
  \bibfield  {author} {\bibinfo {author} {\bibfnamefont {S.}~\bibnamefont
  {Baek}},\ }\href@noop {} {\  (\bibinfo {year} {2017})},\ \Eprint
  {http://arxiv.org/abs/1707.04573} {arXiv:1707.04573 [hep-ph]} \BibitemShut
  {NoStop}%
%%CITATION = ARXIV:1707.04573;%%
\bibitem [{\citenamefont {Ko}\ \emph {et~al.}(2017)\citenamefont {Ko},
  \citenamefont {Nomura},\ and\ \citenamefont {Okada}}]{Ko:2017yrd}%
  \BibitemOpen
  \bibfield  {author} {\bibinfo {author} {\bibfnamefont {P.}~\bibnamefont
  {Ko}}, \bibinfo {author} {\bibfnamefont {T.}~\bibnamefont {Nomura}}, \ and\
  \bibinfo {author} {\bibfnamefont {H.}~\bibnamefont {Okada}},\ }\href
  {\doibase 10.1103/PhysRevD.95.111701} {\bibfield  {journal} {\bibinfo
  {journal} {Phys. Rev.}\ }\textbf {\bibinfo {volume} {D95}},\ \bibinfo {pages}
  {111701} (\bibinfo {year} {2017})},\ \Eprint
  {http://arxiv.org/abs/1702.02699} {arXiv:1702.02699 [hep-ph]} \BibitemShut
  {NoStop}%
%%CITATION = ARXIV:1702.02699;%%
\bibitem [{\citenamefont {Chen}\ and\ \citenamefont
  {Nomura}(2018)}]{Chen:2017usq}%
  \BibitemOpen
  \bibfield  {author} {\bibinfo {author} {\bibfnamefont {C.-H.}\ \bibnamefont
  {Chen}}\ and\ \bibinfo {author} {\bibfnamefont {T.}~\bibnamefont {Nomura}},\
  }\href {\doibase 10.1016/j.physletb.2017.12.062} {\bibfield  {journal}
  {\bibinfo  {journal} {Phys. Lett.}\ }\textbf {\bibinfo {volume} {B777}},\
  \bibinfo {pages} {420} (\bibinfo {year} {2018})},\ \Eprint
  {http://arxiv.org/abs/1707.03249} {arXiv:1707.03249 [hep-ph]} \BibitemShut
  {NoStop}%
%%CITATION = ARXIV:1707.03249;%%
\bibitem [{\citenamefont {Chen}\ and\ \citenamefont
  {Nomura}(2017{\natexlab{a}})}]{Chen:2017cic}%
  \BibitemOpen
  \bibfield  {author} {\bibinfo {author} {\bibfnamefont {C.-H.}\ \bibnamefont
  {Chen}}\ and\ \bibinfo {author} {\bibfnamefont {T.}~\bibnamefont {Nomura}},\
  }\href {\doibase 10.1103/PhysRevD.96.095023} {\bibfield  {journal} {\bibinfo
  {journal} {Phys. Rev.}\ }\textbf {\bibinfo {volume} {D96}},\ \bibinfo {pages}
  {095023} (\bibinfo {year} {2017}{\natexlab{a}})},\ \Eprint
  {http://arxiv.org/abs/1704.04407} {arXiv:1704.04407 [hep-ph]} \BibitemShut
  {NoStop}%
%%CITATION = ARXIV:1704.04407;%%
\bibitem [{\citenamefont {Gninenko}\ and\ \citenamefont
  {Krasnikov}(2018)}]{Gninenko:2018tlp}%
  \BibitemOpen
  \bibfield  {author} {\bibinfo {author} {\bibfnamefont {S.~N.}\ \bibnamefont
  {Gninenko}}\ and\ \bibinfo {author} {\bibfnamefont {N.~V.}\ \bibnamefont
  {Krasnikov}},\ }\href@noop {} {\  (\bibinfo {year} {2018})},\ \Eprint
  {http://arxiv.org/abs/1801.10448} {arXiv:1801.10448 [hep-ph]} \BibitemShut
  {NoStop}%
%%CITATION = ARXIV:1801.10448;%%
\bibitem [{\citenamefont {Bian}\ \emph
  {et~al.}(2017{\natexlab{a}})\citenamefont {Bian}, \citenamefont {Choi},
  \citenamefont {Kang},\ and\ \citenamefont {Lee}}]{Bian:2017rpg}%
  \BibitemOpen
  \bibfield  {author} {\bibinfo {author} {\bibfnamefont {L.}~\bibnamefont
  {Bian}}, \bibinfo {author} {\bibfnamefont {S.-M.}\ \bibnamefont {Choi}},
  \bibinfo {author} {\bibfnamefont {Y.-J.}\ \bibnamefont {Kang}}, \ and\
  \bibinfo {author} {\bibfnamefont {H.~M.}\ \bibnamefont {Lee}},\ }\href
  {\doibase 10.1103/PhysRevD.96.075038} {\bibfield  {journal} {\bibinfo
  {journal} {Phys. Rev.}\ }\textbf {\bibinfo {volume} {D96}},\ \bibinfo {pages}
  {075038} (\bibinfo {year} {2017}{\natexlab{a}})},\ \Eprint
  {http://arxiv.org/abs/1707.04811} {arXiv:1707.04811 [hep-ph]} \BibitemShut
  {NoStop}%
%%CITATION = ARXIV:1707.04811;%%
\bibitem [{\citenamefont {Bian}\ \emph
  {et~al.}(2017{\natexlab{b}})\citenamefont {Bian}, \citenamefont {Lee},\ and\
  \citenamefont {Park}}]{Bian:2017xzg}%
  \BibitemOpen
  \bibfield  {author} {\bibinfo {author} {\bibfnamefont {L.}~\bibnamefont
  {Bian}}, \bibinfo {author} {\bibfnamefont {H.~M.}\ \bibnamefont {Lee}}, \
  and\ \bibinfo {author} {\bibfnamefont {C.~B.}\ \bibnamefont {Park}},\
  }\href@noop {} {\  (\bibinfo {year} {2017}{\natexlab{b}})},\ \Eprint
  {http://arxiv.org/abs/1711.08930} {arXiv:1711.08930 [hep-ph]} \BibitemShut
  {NoStop}%
%%CITATION = ARXIV:1711.08930;%%
\bibitem [{\citenamefont {Gupta}\ \emph {et~al.}(2013)\citenamefont {Gupta},
  \citenamefont {Joshipura},\ and\ \citenamefont {Patel}}]{Gupta:2013it}%
  \BibitemOpen
  \bibfield  {author} {\bibinfo {author} {\bibfnamefont {S.}~\bibnamefont
  {Gupta}}, \bibinfo {author} {\bibfnamefont {A.~S.}\ \bibnamefont
  {Joshipura}}, \ and\ \bibinfo {author} {\bibfnamefont {K.~M.}\ \bibnamefont
  {Patel}},\ }\href {\doibase 10.1007/JHEP09(2013)035} {\bibfield  {journal}
  {\bibinfo  {journal} {JHEP}\ }\textbf {\bibinfo {volume} {09}},\ \bibinfo
  {pages} {035} (\bibinfo {year} {2013})},\ \Eprint
  {http://arxiv.org/abs/1301.7130} {arXiv:1301.7130 [hep-ph]} \BibitemShut
  {NoStop}%
%%CITATION = ARXIV:1301.7130;%%
\bibitem [{\citenamefont {Chen}\ and\ \citenamefont
  {Nomura}(2017{\natexlab{b}})}]{Chen:2017gvf}%
  \BibitemOpen
  \bibfield  {author} {\bibinfo {author} {\bibfnamefont {C.-H.}\ \bibnamefont
  {Chen}}\ and\ \bibinfo {author} {\bibfnamefont {T.}~\bibnamefont {Nomura}},\
  }\href@noop {} {\  (\bibinfo {year} {2017}{\natexlab{b}})},\ \Eprint
  {http://arxiv.org/abs/1705.10620} {arXiv:1705.10620 [hep-ph]} \BibitemShut
  {NoStop}%
%%CITATION = ARXIV:1705.10620;%%
\bibitem [{\citenamefont {Dev}(2017)}]{Dev:2017fdz}%
  \BibitemOpen
  \bibfield  {author} {\bibinfo {author} {\bibfnamefont {A.}~\bibnamefont
  {Dev}},\ }\href@noop {} {\  (\bibinfo {year} {2017})},\ \Eprint
  {http://arxiv.org/abs/1710.02878} {arXiv:1710.02878 [hep-ph]} \BibitemShut
  {NoStop}%
%%CITATION = ARXIV:1710.02878;%%
\bibitem [{\citenamefont {Nath}\ \emph {et~al.}(2018)\citenamefont {Nath},
  \citenamefont {Xing},\ and\ \citenamefont {Zhang}}]{Nath:2018hjx}%
  \BibitemOpen
  \bibfield  {author} {\bibinfo {author} {\bibfnamefont {N.}~\bibnamefont
  {Nath}}, \bibinfo {author} {\bibfnamefont {Z.-z.}\ \bibnamefont {Xing}}, \
  and\ \bibinfo {author} {\bibfnamefont {J.}~\bibnamefont {Zhang}},\
  }\href@noop {} {\  (\bibinfo {year} {2018})},\ \Eprint
  {http://arxiv.org/abs/1801.09931} {arXiv:1801.09931 [hep-ph]} \BibitemShut
  {NoStop}%
%%CITATION = ARXIV:1801.09931;%%
\bibitem [{\citenamefont {Elahi}\ and\ \citenamefont
  {Martin}(2016)}]{Elahi:2015vzh}%
  \BibitemOpen
  \bibfield  {author} {\bibinfo {author} {\bibfnamefont {F.}~\bibnamefont
  {Elahi}}\ and\ \bibinfo {author} {\bibfnamefont {A.}~\bibnamefont {Martin}},\
  }\href {\doibase 10.1103/PhysRevD.93.015022} {\bibfield  {journal} {\bibinfo
  {journal} {Phys. Rev.}\ }\textbf {\bibinfo {volume} {D93}},\ \bibinfo {pages}
  {015022} (\bibinfo {year} {2016})},\ \Eprint
  {http://arxiv.org/abs/1511.04107} {arXiv:1511.04107 [hep-ph]} \BibitemShut
  {NoStop}%
%%CITATION = ARXIV:1511.04107;%%
\bibitem [{\citenamefont {Kaneta}\ and\ \citenamefont
  {Shimomura}(2017)}]{Kaneta:2016uyt}%
  \BibitemOpen
  \bibfield  {author} {\bibinfo {author} {\bibfnamefont {Y.}~\bibnamefont
  {Kaneta}}\ and\ \bibinfo {author} {\bibfnamefont {T.}~\bibnamefont
  {Shimomura}},\ }\href {\doibase 10.1093/ptep/ptx050} {\bibfield  {journal}
  {\bibinfo  {journal} {PTEP}\ }\textbf {\bibinfo {volume} {2017}},\ \bibinfo
  {pages} {053B04} (\bibinfo {year} {2017})},\ \Eprint
  {http://arxiv.org/abs/1701.00156} {arXiv:1701.00156 [hep-ph]} \BibitemShut
  {NoStop}%
%%CITATION = ARXIV:1701.00156;%%
\bibitem [{\citenamefont {Araki}\ \emph {et~al.}(2017)\citenamefont {Araki},
  \citenamefont {Hoshino}, \citenamefont {Ota}, \citenamefont {Sato},\ and\
  \citenamefont {Shimomura}}]{Araki:2017wyg}%
  \BibitemOpen
  \bibfield  {author} {\bibinfo {author} {\bibfnamefont {T.}~\bibnamefont
  {Araki}}, \bibinfo {author} {\bibfnamefont {S.}~\bibnamefont {Hoshino}},
  \bibinfo {author} {\bibfnamefont {T.}~\bibnamefont {Ota}}, \bibinfo {author}
  {\bibfnamefont {J.}~\bibnamefont {Sato}}, \ and\ \bibinfo {author}
  {\bibfnamefont {T.}~\bibnamefont {Shimomura}},\ }\href {\doibase
  10.1103/PhysRevD.95.055006} {\bibfield  {journal} {\bibinfo  {journal} {Phys.
  Rev.}\ }\textbf {\bibinfo {volume} {D95}},\ \bibinfo {pages} {055006}
  (\bibinfo {year} {2017})},\ \Eprint {http://arxiv.org/abs/1702.01497}
  {arXiv:1702.01497 [hep-ph]} \BibitemShut {NoStop}%
%%CITATION = ARXIV:1702.01497;%%
\bibitem [{\citenamefont {Nomura}\ and\ \citenamefont
  {Shimomura}(2018)}]{Nomura:2018yej}%
  \BibitemOpen
  \bibfield  {author} {\bibinfo {author} {\bibfnamefont {T.}~\bibnamefont
  {Nomura}}\ and\ \bibinfo {author} {\bibfnamefont {T.}~\bibnamefont
  {Shimomura}},\ }\href@noop {} {\  (\bibinfo {year} {2018})},\ \Eprint
  {http://arxiv.org/abs/1803.00842} {arXiv:1803.00842 [hep-ph]} \BibitemShut
  {NoStop}%
%%CITATION = ARXIV:1803.00842;%%
\bibitem [{\citenamefont {Park}\ \emph {et~al.}(2016)\citenamefont {Park},
  \citenamefont {Kim},\ and\ \citenamefont {Park}}]{Kim:2015fpa}%
  \BibitemOpen
  \bibfield  {author} {\bibinfo {author} {\bibfnamefont {J.-C.}\ \bibnamefont
  {Park}}, \bibinfo {author} {\bibfnamefont {J.}~\bibnamefont {Kim}}, \ and\
  \bibinfo {author} {\bibfnamefont {S.~C.}\ \bibnamefont {Park}},\ }\href
  {\doibase 10.1016/j.physletb.2015.11.035} {\bibfield  {journal} {\bibinfo
  {journal} {Phys. Lett.}\ }\textbf {\bibinfo {volume} {B752}},\ \bibinfo
  {pages} {59} (\bibinfo {year} {2016})},\ \Eprint
  {http://arxiv.org/abs/1505.04620} {arXiv:1505.04620 [hep-ph]} \BibitemShut
  {NoStop}%
%%CITATION = ARXIV:1505.04620;%%
\bibitem [{\citenamefont {Kohda}\ \emph {et~al.}(2018)\citenamefont {Kohda},
  \citenamefont {Modak},\ and\ \citenamefont {Soffer}}]{Kohda:2018xbc}%
  \BibitemOpen
  \bibfield  {author} {\bibinfo {author} {\bibfnamefont {M.}~\bibnamefont
  {Kohda}}, \bibinfo {author} {\bibfnamefont {T.}~\bibnamefont {Modak}}, \ and\
  \bibinfo {author} {\bibfnamefont {A.}~\bibnamefont {Soffer}},\ }\href@noop {}
  {\  (\bibinfo {year} {2018})},\ \Eprint {http://arxiv.org/abs/1803.07492}
  {arXiv:1803.07492 [hep-ph]} \BibitemShut {NoStop}%
%%CITATION = ARXIV:1803.07492;%%
\bibitem [{\citenamefont {Hou}\ \emph {et~al.}(2018)\citenamefont {Hou},
  \citenamefont {Kohda},\ and\ \citenamefont {Modak}}]{Hou:2018npi}%
  \BibitemOpen
  \bibfield  {author} {\bibinfo {author} {\bibfnamefont {W.-S.}\ \bibnamefont
  {Hou}}, \bibinfo {author} {\bibfnamefont {M.}~\bibnamefont {Kohda}}, \ and\
  \bibinfo {author} {\bibfnamefont {T.}~\bibnamefont {Modak}},\ }\href@noop {}
  {\  (\bibinfo {year} {2018})},\ \Eprint {http://arxiv.org/abs/1801.02579}
  {arXiv:1801.02579 [hep-ph]} \BibitemShut {NoStop}%
%%CITATION = ARXIV:1801.02579;%%
\bibitem [{\citenamefont {Biswas}\ \emph {et~al.}(2017)\citenamefont {Biswas},
  \citenamefont {Choubey},\ and\ \citenamefont {Khan}}]{Biswas:2016yjr}%
  \BibitemOpen
  \bibfield  {author} {\bibinfo {author} {\bibfnamefont {A.}~\bibnamefont
  {Biswas}}, \bibinfo {author} {\bibfnamefont {S.}~\bibnamefont {Choubey}}, \
  and\ \bibinfo {author} {\bibfnamefont {S.}~\bibnamefont {Khan}},\ }\href
  {\doibase 10.1007/JHEP02(2017)123} {\bibfield  {journal} {\bibinfo  {journal}
  {JHEP}\ }\textbf {\bibinfo {volume} {02}},\ \bibinfo {pages} {123} (\bibinfo
  {year} {2017})},\ \Eprint {http://arxiv.org/abs/1612.03067} {arXiv:1612.03067
  [hep-ph]} \BibitemShut {NoStop}%
%%CITATION = ARXIV:1612.03067;%%
\bibitem [{\citenamefont {Baek}\ and\ \citenamefont {Ko}(2009)}]{Baek:2008nz}%
  \BibitemOpen
  \bibfield  {author} {\bibinfo {author} {\bibfnamefont {S.}~\bibnamefont
  {Baek}}\ and\ \bibinfo {author} {\bibfnamefont {P.}~\bibnamefont {Ko}},\
  }\href {\doibase 10.1088/1475-7516/2009/10/011} {\bibfield  {journal}
  {\bibinfo  {journal} {JCAP}\ }\textbf {\bibinfo {volume} {0910}},\ \bibinfo
  {pages} {011} (\bibinfo {year} {2009})},\ \Eprint
  {http://arxiv.org/abs/0811.1646} {arXiv:0811.1646 [hep-ph]} \BibitemShut
  {NoStop}%
%%CITATION = ARXIV:0811.1646;%%
\bibitem [{\citenamefont {Altmannshofer}\ \emph
  {et~al.}(2016{\natexlab{b}})\citenamefont {Altmannshofer}, \citenamefont
  {Gori}, \citenamefont {Profumo},\ and\ \citenamefont
  {Queiroz}}]{Altmannshofer:2016jzy}%
  \BibitemOpen
  \bibfield  {author} {\bibinfo {author} {\bibfnamefont {W.}~\bibnamefont
  {Altmannshofer}}, \bibinfo {author} {\bibfnamefont {S.}~\bibnamefont {Gori}},
  \bibinfo {author} {\bibfnamefont {S.}~\bibnamefont {Profumo}}, \ and\
  \bibinfo {author} {\bibfnamefont {F.~S.}\ \bibnamefont {Queiroz}},\ }\href
  {\doibase 10.1007/JHEP12(2016)106} {\bibfield  {journal} {\bibinfo  {journal}
  {JHEP}\ }\textbf {\bibinfo {volume} {12}},\ \bibinfo {pages} {106} (\bibinfo
  {year} {2016}{\natexlab{b}})},\ \Eprint {http://arxiv.org/abs/1609.04026}
  {arXiv:1609.04026 [hep-ph]} \BibitemShut {NoStop}%
%%CITATION = ARXIV:1609.04026;%%
\bibitem [{\citenamefont {Baek}(2016)}]{Baek:2015fea}%
  \BibitemOpen
  \bibfield  {author} {\bibinfo {author} {\bibfnamefont {S.}~\bibnamefont
  {Baek}},\ }\href {\doibase 10.1016/j.physletb.2016.02.062} {\bibfield
  {journal} {\bibinfo  {journal} {Phys. Lett.}\ }\textbf {\bibinfo {volume}
  {B756}},\ \bibinfo {pages} {1} (\bibinfo {year} {2016})},\ \Eprint
  {http://arxiv.org/abs/1510.02168} {arXiv:1510.02168 [hep-ph]} \BibitemShut
  {NoStop}%
%%CITATION = ARXIV:1510.02168;%%
\bibitem [{\citenamefont {Biswas}\ \emph {et~al.}(2016)\citenamefont {Biswas},
  \citenamefont {Choubey},\ and\ \citenamefont {Khan}}]{Biswas:2016yan}%
  \BibitemOpen
  \bibfield  {author} {\bibinfo {author} {\bibfnamefont {A.}~\bibnamefont
  {Biswas}}, \bibinfo {author} {\bibfnamefont {S.}~\bibnamefont {Choubey}}, \
  and\ \bibinfo {author} {\bibfnamefont {S.}~\bibnamefont {Khan}},\ }\href
  {\doibase 10.1007/JHEP09(2016)147} {\bibfield  {journal} {\bibinfo  {journal}
  {JHEP}\ }\textbf {\bibinfo {volume} {09}},\ \bibinfo {pages} {147} (\bibinfo
  {year} {2016})},\ \Eprint {http://arxiv.org/abs/1608.04194} {arXiv:1608.04194
  [hep-ph]} \BibitemShut {NoStop}%
%%CITATION = ARXIV:1608.04194;%%
\bibitem [{\citenamefont {Patra}\ \emph {et~al.}(2017)\citenamefont {Patra},
  \citenamefont {Rao}, \citenamefont {Sahoo},\ and\ \citenamefont
  {Sahu}}]{Patra:2016shz}%
  \BibitemOpen
  \bibfield  {author} {\bibinfo {author} {\bibfnamefont {S.}~\bibnamefont
  {Patra}}, \bibinfo {author} {\bibfnamefont {S.}~\bibnamefont {Rao}}, \bibinfo
  {author} {\bibfnamefont {N.}~\bibnamefont {Sahoo}}, \ and\ \bibinfo {author}
  {\bibfnamefont {N.}~\bibnamefont {Sahu}},\ }\href {\doibase
  10.1016/j.nuclphysb.2017.02.010} {\bibfield  {journal} {\bibinfo  {journal}
  {Nucl. Phys.}\ }\textbf {\bibinfo {volume} {B917}},\ \bibinfo {pages} {317}
  (\bibinfo {year} {2017})},\ \Eprint {http://arxiv.org/abs/1607.04046}
  {arXiv:1607.04046 [hep-ph]} \BibitemShut {NoStop}%
%%CITATION = ARXIV:1607.04046;%%
\bibitem [{\citenamefont {Holdom}(1986)}]{Holdom:1985ag}%
  \BibitemOpen
  \bibfield  {author} {\bibinfo {author} {\bibfnamefont {B.}~\bibnamefont
  {Holdom}},\ }\href {\doibase 10.1016/0370-2693(86)91377-8} {\bibfield
  {journal} {\bibinfo  {journal} {Phys. Lett.}\ }\textbf {\bibinfo {volume}
  {166B}},\ \bibinfo {pages} {196} (\bibinfo {year} {1986})}\BibitemShut
  {NoStop}%
%%CITATION = PHLTA,166B,196;%%
\bibitem [{\citenamefont {Carone}\ and\ \citenamefont
  {Murayama}(1995)}]{Carone:1995pu}%
  \BibitemOpen
  \bibfield  {author} {\bibinfo {author} {\bibfnamefont {C.~D.}\ \bibnamefont
  {Carone}}\ and\ \bibinfo {author} {\bibfnamefont {H.}~\bibnamefont
  {Murayama}},\ }\href {\doibase 10.1103/PhysRevD.52.484} {\bibfield  {journal}
  {\bibinfo  {journal} {Phys. Rev.}\ }\textbf {\bibinfo {volume} {D52}},\
  \bibinfo {pages} {484} (\bibinfo {year} {1995})},\ \Eprint
  {http://arxiv.org/abs/hep-ph/9501220} {arXiv:hep-ph/9501220 [hep-ph]}
  \BibitemShut {NoStop}%
%%CITATION = HEP-PH/9501220;%%
\bibitem [{\citenamefont {Kahlhoefer}\ \emph {et~al.}(2016)\citenamefont
  {Kahlhoefer}, \citenamefont {Schmidt-Hoberg}, \citenamefont {Schwetz},\ and\
  \citenamefont {Vogl}}]{Kahlhoefer:2015bea}%
  \BibitemOpen
  \bibfield  {author} {\bibinfo {author} {\bibfnamefont {F.}~\bibnamefont
  {Kahlhoefer}}, \bibinfo {author} {\bibfnamefont {K.}~\bibnamefont
  {Schmidt-Hoberg}}, \bibinfo {author} {\bibfnamefont {T.}~\bibnamefont
  {Schwetz}}, \ and\ \bibinfo {author} {\bibfnamefont {S.}~\bibnamefont
  {Vogl}},\ }\href {\doibase 10.1007/JHEP02(2016)016} {\bibfield  {journal}
  {\bibinfo  {journal} {JHEP}\ }\textbf {\bibinfo {volume} {02}},\ \bibinfo
  {pages} {016} (\bibinfo {year} {2016})},\ \Eprint
  {http://arxiv.org/abs/1510.02110} {arXiv:1510.02110 [hep-ph]} \BibitemShut
  {NoStop}%
%%CITATION = ARXIV:1510.02110;%%
\bibitem [{\citenamefont {Descotes-Genon}\ \emph {et~al.}(2013)\citenamefont
  {Descotes-Genon}, \citenamefont {Matias},\ and\ \citenamefont
  {Virto}}]{Descotes-Genon:2013wba}%
  \BibitemOpen
  \bibfield  {author} {\bibinfo {author} {\bibfnamefont {S.}~\bibnamefont
  {Descotes-Genon}}, \bibinfo {author} {\bibfnamefont {J.}~\bibnamefont
  {Matias}}, \ and\ \bibinfo {author} {\bibfnamefont {J.}~\bibnamefont
  {Virto}},\ }\href {\doibase 10.1103/PhysRevD.88.074002} {\bibfield  {journal}
  {\bibinfo  {journal} {Phys. Rev.}\ }\textbf {\bibinfo {volume} {D88}},\
  \bibinfo {pages} {074002} (\bibinfo {year} {2013})},\ \Eprint
  {http://arxiv.org/abs/1307.5683} {arXiv:1307.5683 [hep-ph]} \BibitemShut
  {NoStop}%
%%CITATION = ARXIV:1307.5683;%%
\bibitem [{\citenamefont {Altmannshofer}\ and\ \citenamefont
  {Straub}(2013)}]{Altmannshofer:2013foa}%
  \BibitemOpen
  \bibfield  {author} {\bibinfo {author} {\bibfnamefont {W.}~\bibnamefont
  {Altmannshofer}}\ and\ \bibinfo {author} {\bibfnamefont {D.~M.}\ \bibnamefont
  {Straub}},\ }\href {\doibase 10.1140/epjc/s10052-013-2646-9} {\bibfield
  {journal} {\bibinfo  {journal} {Eur. Phys. J.}\ }\textbf {\bibinfo {volume}
  {C73}},\ \bibinfo {pages} {2646} (\bibinfo {year} {2013})},\ \Eprint
  {http://arxiv.org/abs/1308.1501} {arXiv:1308.1501 [hep-ph]} \BibitemShut
  {NoStop}%
%%CITATION = ARXIV:1308.1501;%%
\bibitem [{\citenamefont {Beaujean}\ \emph {et~al.}(2014)\citenamefont
  {Beaujean}, \citenamefont {Bobeth},\ and\ \citenamefont {van
  Dyk}}]{Beaujean:2013soa}%
  \BibitemOpen
  \bibfield  {author} {\bibinfo {author} {\bibfnamefont {F.}~\bibnamefont
  {Beaujean}}, \bibinfo {author} {\bibfnamefont {C.}~\bibnamefont {Bobeth}}, \
  and\ \bibinfo {author} {\bibfnamefont {D.}~\bibnamefont {van Dyk}},\ }\href
  {\doibase 10.1140/epjc/s10052-014-2897-0, 10.1140/epjc/s10052-014-3179-6}
  {\bibfield  {journal} {\bibinfo  {journal} {Eur. Phys. J.}\ }\textbf
  {\bibinfo {volume} {C74}},\ \bibinfo {pages} {2897} (\bibinfo {year}
  {2014})},\ \bibinfo {note} {[Erratum: Eur. Phys. J.C74,3179(2014)]},\ \Eprint
  {http://arxiv.org/abs/1310.2478} {arXiv:1310.2478 [hep-ph]} \BibitemShut
  {NoStop}%
%%CITATION = ARXIV:1310.2478;%%
\bibitem [{\citenamefont {Hurth}\ and\ \citenamefont
  {Mahmoudi}(2014)}]{Hurth:2013ssa}%
  \BibitemOpen
  \bibfield  {author} {\bibinfo {author} {\bibfnamefont {T.}~\bibnamefont
  {Hurth}}\ and\ \bibinfo {author} {\bibfnamefont {F.}~\bibnamefont
  {Mahmoudi}},\ }\href {\doibase 10.1007/JHEP04(2014)097} {\bibfield  {journal}
  {\bibinfo  {journal} {JHEP}\ }\textbf {\bibinfo {volume} {04}},\ \bibinfo
  {pages} {097} (\bibinfo {year} {2014})},\ \Eprint
  {http://arxiv.org/abs/1312.5267} {arXiv:1312.5267 [hep-ph]} \BibitemShut
  {NoStop}%
%%CITATION = ARXIV:1312.5267;%%
\bibitem [{\citenamefont {Aaij}\ \emph {et~al.}(2016)\citenamefont {Aaij} \emph
  {et~al.}}]{Aaij:2015oid}%
  \BibitemOpen
  \bibfield  {author} {\bibinfo {author} {\bibfnamefont {R.}~\bibnamefont
  {Aaij}} \emph {et~al.} (\bibinfo {collaboration} {LHCb}),\ }\href {\doibase
  10.1007/JHEP02(2016)104} {\bibfield  {journal} {\bibinfo  {journal} {JHEP}\
  }\textbf {\bibinfo {volume} {02}},\ \bibinfo {pages} {104} (\bibinfo {year}
  {2016})},\ \Eprint {http://arxiv.org/abs/1512.04442} {arXiv:1512.04442
  [hep-ex]} \BibitemShut {NoStop}%
%%CITATION = ARXIV:1512.04442;%%
\bibitem [{\citenamefont {Aaij}\ \emph {et~al.}(2014)\citenamefont {Aaij} \emph
  {et~al.}}]{Aaij:2014ora}%
  \BibitemOpen
  \bibfield  {author} {\bibinfo {author} {\bibfnamefont {R.}~\bibnamefont
  {Aaij}} \emph {et~al.} (\bibinfo {collaboration} {LHCb}),\ }\href {\doibase
  10.1103/PhysRevLett.113.151601} {\bibfield  {journal} {\bibinfo  {journal}
  {Phys. Rev. Lett.}\ }\textbf {\bibinfo {volume} {113}},\ \bibinfo {pages}
  {151601} (\bibinfo {year} {2014})},\ \Eprint {http://arxiv.org/abs/1406.6482}
  {arXiv:1406.6482 [hep-ex]} \BibitemShut {NoStop}%
%%CITATION = ARXIV:1406.6482;%%
\bibitem [{\citenamefont {Altmannshofer}\ \emph {et~al.}(2012)\citenamefont
  {Altmannshofer}, \citenamefont {Paradisi},\ and\ \citenamefont
  {Straub}}]{Altmannshofer:2011gn}%
  \BibitemOpen
  \bibfield  {author} {\bibinfo {author} {\bibfnamefont {W.}~\bibnamefont
  {Altmannshofer}}, \bibinfo {author} {\bibfnamefont {P.}~\bibnamefont
  {Paradisi}}, \ and\ \bibinfo {author} {\bibfnamefont {D.~M.}\ \bibnamefont
  {Straub}},\ }\href {\doibase 10.1007/JHEP04(2012)008} {\bibfield  {journal}
  {\bibinfo  {journal} {JHEP}\ }\textbf {\bibinfo {volume} {04}},\ \bibinfo
  {pages} {008} (\bibinfo {year} {2012})},\ \Eprint
  {http://arxiv.org/abs/1111.1257} {arXiv:1111.1257 [hep-ph]} \BibitemShut
  {NoStop}%
%%CITATION = ARXIV:1111.1257;%%
\bibitem [{\citenamefont {Altmannshofer}\ and\ \citenamefont
  {Straub}(2012)}]{Altmannshofer:2012az}%
  \BibitemOpen
  \bibfield  {author} {\bibinfo {author} {\bibfnamefont {W.}~\bibnamefont
  {Altmannshofer}}\ and\ \bibinfo {author} {\bibfnamefont {D.~M.}\ \bibnamefont
  {Straub}},\ }\href {\doibase 10.1007/JHEP08(2012)121} {\bibfield  {journal}
  {\bibinfo  {journal} {JHEP}\ }\textbf {\bibinfo {volume} {08}},\ \bibinfo
  {pages} {121} (\bibinfo {year} {2012})},\ \Eprint
  {http://arxiv.org/abs/1206.0273} {arXiv:1206.0273 [hep-ph]} \BibitemShut
  {NoStop}%
%%CITATION = ARXIV:1206.0273;%%
\bibitem [{\citenamefont {Altmannshofer}\ and\ \citenamefont
  {Straub}(2015)}]{Altmannshofer:2014rta}%
  \BibitemOpen
  \bibfield  {author} {\bibinfo {author} {\bibfnamefont {W.}~\bibnamefont
  {Altmannshofer}}\ and\ \bibinfo {author} {\bibfnamefont {D.~M.}\ \bibnamefont
  {Straub}},\ }\href {\doibase 10.1140/epjc/s10052-015-3602-7} {\bibfield
  {journal} {\bibinfo  {journal} {Eur. Phys. J.}\ }\textbf {\bibinfo {volume}
  {C75}},\ \bibinfo {pages} {382} (\bibinfo {year} {2015})},\ \Eprint
  {http://arxiv.org/abs/1411.3161} {arXiv:1411.3161 [hep-ph]} \BibitemShut
  {NoStop}%
%%CITATION = ARXIV:1411.3161;%%
\bibitem [{\citenamefont {Descotes-Genon}\ \emph {et~al.}(2016)\citenamefont
  {Descotes-Genon}, \citenamefont {Hofer}, \citenamefont {Matias},\ and\
  \citenamefont {Virto}}]{Descotes-Genon:2015uva}%
  \BibitemOpen
  \bibfield  {author} {\bibinfo {author} {\bibfnamefont {S.}~\bibnamefont
  {Descotes-Genon}}, \bibinfo {author} {\bibfnamefont {L.}~\bibnamefont
  {Hofer}}, \bibinfo {author} {\bibfnamefont {J.}~\bibnamefont {Matias}}, \
  and\ \bibinfo {author} {\bibfnamefont {J.}~\bibnamefont {Virto}},\ }\href
  {\doibase 10.1007/JHEP06(2016)092} {\bibfield  {journal} {\bibinfo  {journal}
  {JHEP}\ }\textbf {\bibinfo {volume} {06}},\ \bibinfo {pages} {092} (\bibinfo
  {year} {2016})},\ \Eprint {http://arxiv.org/abs/1510.04239} {arXiv:1510.04239
  [hep-ph]} \BibitemShut {NoStop}%
%%CITATION = ARXIV:1510.04239;%%
\bibitem [{\citenamefont {Neshatpour}\ \emph {et~al.}(2017)\citenamefont
  {Neshatpour}, \citenamefont {Chobanova}, \citenamefont {Hurth}, \citenamefont
  {Mahmoudi},\ and\ \citenamefont {Martinez~Santos}}]{Neshatpour:2017qvi}%
  \BibitemOpen
  \bibfield  {author} {\bibinfo {author} {\bibfnamefont {S.}~\bibnamefont
  {Neshatpour}}, \bibinfo {author} {\bibfnamefont {V.~G.}\ \bibnamefont
  {Chobanova}}, \bibinfo {author} {\bibfnamefont {T.}~\bibnamefont {Hurth}},
  \bibinfo {author} {\bibfnamefont {F.}~\bibnamefont {Mahmoudi}}, \ and\
  \bibinfo {author} {\bibfnamefont {D.}~\bibnamefont {Martinez~Santos}},\ }in\
  \href {http://inspirehep.net/record/1601662/files/arXiv:1705.10730.pdf}
  {\emph {\bibinfo {booktitle} {{Proceedings, 52nd Rencontres de Moriond on QCD
  and High Energy Interactions: La Thuile, Italy, March 25-April 1, 2017}}}}\
  (\bibinfo {year} {2017})\ \Eprint {http://arxiv.org/abs/1705.10730}
  {arXiv:1705.10730 [hep-ph]} \BibitemShut {NoStop}%
%%CITATION = ARXIV:1705.10730;%%
\bibitem [{\citenamefont {Capdevila}\ \emph {et~al.}(2018)\citenamefont
  {Capdevila}, \citenamefont {Crivellin}, \citenamefont {Descotes-Genon},
  \citenamefont {Matias},\ and\ \citenamefont {Virto}}]{Capdevila:2017bsm}%
  \BibitemOpen
  \bibfield  {author} {\bibinfo {author} {\bibfnamefont {B.}~\bibnamefont
  {Capdevila}}, \bibinfo {author} {\bibfnamefont {A.}~\bibnamefont
  {Crivellin}}, \bibinfo {author} {\bibfnamefont {S.}~\bibnamefont
  {Descotes-Genon}}, \bibinfo {author} {\bibfnamefont {J.}~\bibnamefont
  {Matias}}, \ and\ \bibinfo {author} {\bibfnamefont {J.}~\bibnamefont
  {Virto}},\ }\href {\doibase 10.1007/JHEP01(2018)093} {\bibfield  {journal}
  {\bibinfo  {journal} {JHEP}\ }\textbf {\bibinfo {volume} {01}},\ \bibinfo
  {pages} {093} (\bibinfo {year} {2018})},\ \Eprint
  {http://arxiv.org/abs/1704.05340} {arXiv:1704.05340 [hep-ph]} \BibitemShut
  {NoStop}%
%%CITATION = ARXIV:1704.05340;%%
\bibitem [{\citenamefont {Altmannshofer}\ \emph {et~al.}(2017)\citenamefont
  {Altmannshofer}, \citenamefont {Niehoff}, \citenamefont {Stangl},\ and\
  \citenamefont {Straub}}]{Altmannshofer:2017fio}%
  \BibitemOpen
  \bibfield  {author} {\bibinfo {author} {\bibfnamefont {W.}~\bibnamefont
  {Altmannshofer}}, \bibinfo {author} {\bibfnamefont {C.}~\bibnamefont
  {Niehoff}}, \bibinfo {author} {\bibfnamefont {P.}~\bibnamefont {Stangl}}, \
  and\ \bibinfo {author} {\bibfnamefont {D.~M.}\ \bibnamefont {Straub}},\
  }\href {\doibase 10.1140/epjc/s10052-017-4952-0} {\bibfield  {journal}
  {\bibinfo  {journal} {Eur. Phys. J.}\ }\textbf {\bibinfo {volume} {C77}},\
  \bibinfo {pages} {377} (\bibinfo {year} {2017})},\ \Eprint
  {http://arxiv.org/abs/1703.09189} {arXiv:1703.09189 [hep-ph]} \BibitemShut
  {NoStop}%
%%CITATION = ARXIV:1703.09189;%%
\bibitem [{\citenamefont {Ciuchini}\ \emph {et~al.}(2017)\citenamefont
  {Ciuchini}, \citenamefont {Coutinho}, \citenamefont {Fedele}, \citenamefont
  {Franco}, \citenamefont {Paul}, \citenamefont {Silvestrini},\ and\
  \citenamefont {Valli}}]{Ciuchini:2017mik}%
  \BibitemOpen
  \bibfield  {author} {\bibinfo {author} {\bibfnamefont {M.}~\bibnamefont
  {Ciuchini}}, \bibinfo {author} {\bibfnamefont {A.~M.}\ \bibnamefont
  {Coutinho}}, \bibinfo {author} {\bibfnamefont {M.}~\bibnamefont {Fedele}},
  \bibinfo {author} {\bibfnamefont {E.}~\bibnamefont {Franco}}, \bibinfo
  {author} {\bibfnamefont {A.}~\bibnamefont {Paul}}, \bibinfo {author}
  {\bibfnamefont {L.}~\bibnamefont {Silvestrini}}, \ and\ \bibinfo {author}
  {\bibfnamefont {M.}~\bibnamefont {Valli}},\ }\href {\doibase
  10.1140/epjc/s10052-017-5270-2} {\bibfield  {journal} {\bibinfo  {journal}
  {Eur. Phys. J.}\ }\textbf {\bibinfo {volume} {C77}},\ \bibinfo {pages} {688}
  (\bibinfo {year} {2017})},\ \Eprint {http://arxiv.org/abs/1704.05447}
  {arXiv:1704.05447 [hep-ph]} \BibitemShut {NoStop}%
%%CITATION = ARXIV:1704.05447;%%
\bibitem [{\citenamefont {Aristizabal~Sierra}\ \emph
  {et~al.}(2015)\citenamefont {Aristizabal~Sierra}, \citenamefont {Staub},\
  and\ \citenamefont {Vicente}}]{Sierra:2015fma}%
  \BibitemOpen
  \bibfield  {author} {\bibinfo {author} {\bibfnamefont {D.}~\bibnamefont
  {Aristizabal~Sierra}}, \bibinfo {author} {\bibfnamefont {F.}~\bibnamefont
  {Staub}}, \ and\ \bibinfo {author} {\bibfnamefont {A.}~\bibnamefont
  {Vicente}},\ }\href {\doibase 10.1103/PhysRevD.92.015001} {\bibfield
  {journal} {\bibinfo  {journal} {Phys. Rev.}\ }\textbf {\bibinfo {volume}
  {D92}},\ \bibinfo {pages} {015001} (\bibinfo {year} {2015})},\ \Eprint
  {http://arxiv.org/abs/1503.06077} {arXiv:1503.06077 [hep-ph]} \BibitemShut
  {NoStop}%
%%CITATION = ARXIV:1503.06077;%%
\bibitem [{\citenamefont {Kawamura}\ \emph {et~al.}(2017)\citenamefont
  {Kawamura}, \citenamefont {Okawa},\ and\ \citenamefont
  {Omura}}]{Kawamura:2017ecz}%
  \BibitemOpen
  \bibfield  {author} {\bibinfo {author} {\bibfnamefont {J.}~\bibnamefont
  {Kawamura}}, \bibinfo {author} {\bibfnamefont {S.}~\bibnamefont {Okawa}}, \
  and\ \bibinfo {author} {\bibfnamefont {Y.}~\bibnamefont {Omura}},\ }\href
  {\doibase 10.1103/PhysRevD.96.075041} {\bibfield  {journal} {\bibinfo
  {journal} {Phys. Rev.}\ }\textbf {\bibinfo {volume} {D96}},\ \bibinfo {pages}
  {075041} (\bibinfo {year} {2017})},\ \Eprint
  {http://arxiv.org/abs/1706.04344} {arXiv:1706.04344 [hep-ph]} \BibitemShut
  {NoStop}%
%%CITATION = ARXIV:1706.04344;%%
\bibitem [{\citenamefont {Cline}(2018)}]{Cline:2017aed}%
  \BibitemOpen
  \bibfield  {author} {\bibinfo {author} {\bibfnamefont {J.~M.}\ \bibnamefont
  {Cline}},\ }\href {\doibase 10.1103/PhysRevD.97.015013} {\bibfield  {journal}
  {\bibinfo  {journal} {Phys. Rev.}\ }\textbf {\bibinfo {volume} {D97}},\
  \bibinfo {pages} {015013} (\bibinfo {year} {2018})},\ \Eprint
  {http://arxiv.org/abs/1710.02140} {arXiv:1710.02140 [hep-ph]} \BibitemShut
  {NoStop}%
%%CITATION = ARXIV:1710.02140;%%
\bibitem [{\citenamefont {Cline}\ \emph {et~al.}(2017)\citenamefont {Cline},
  \citenamefont {Cornell}, \citenamefont {London},\ and\ \citenamefont
  {Watanabe}}]{Cline:2017lvv}%
  \BibitemOpen
  \bibfield  {author} {\bibinfo {author} {\bibfnamefont {J.~M.}\ \bibnamefont
  {Cline}}, \bibinfo {author} {\bibfnamefont {J.~M.}\ \bibnamefont {Cornell}},
  \bibinfo {author} {\bibfnamefont {D.}~\bibnamefont {London}}, \ and\ \bibinfo
  {author} {\bibfnamefont {R.}~\bibnamefont {Watanabe}},\ }\href {\doibase
  10.1103/PhysRevD.95.095015} {\bibfield  {journal} {\bibinfo  {journal} {Phys.
  Rev.}\ }\textbf {\bibinfo {volume} {D95}},\ \bibinfo {pages} {095015}
  (\bibinfo {year} {2017})},\ \Eprint {http://arxiv.org/abs/1702.00395}
  {arXiv:1702.00395 [hep-ph]} \BibitemShut {NoStop}%
%%CITATION = ARXIV:1702.00395;%%
\bibitem [{\citenamefont {Altmannshofer}\ \emph
  {et~al.}(2014{\natexlab{a}})\citenamefont {Altmannshofer}, \citenamefont
  {Gori}, \citenamefont {Pospelov},\ and\ \citenamefont
  {Yavin}}]{Altmannshofer:2014cfa}%
  \BibitemOpen
  \bibfield  {author} {\bibinfo {author} {\bibfnamefont {W.}~\bibnamefont
  {Altmannshofer}}, \bibinfo {author} {\bibfnamefont {S.}~\bibnamefont {Gori}},
  \bibinfo {author} {\bibfnamefont {M.}~\bibnamefont {Pospelov}}, \ and\
  \bibinfo {author} {\bibfnamefont {I.}~\bibnamefont {Yavin}},\ }\href
  {\doibase 10.1103/PhysRevD.89.095033} {\bibfield  {journal} {\bibinfo
  {journal} {Phys. Rev.}\ }\textbf {\bibinfo {volume} {D89}},\ \bibinfo {pages}
  {095033} (\bibinfo {year} {2014}{\natexlab{a}})},\ \Eprint
  {http://arxiv.org/abs/1403.1269} {arXiv:1403.1269 [hep-ph]} \BibitemShut
  {NoStop}%
%%CITATION = ARXIV:1403.1269;%%
\bibitem [{\citenamefont {Babu}\ \emph {et~al.}(1998)\citenamefont {Babu},
  \citenamefont {Kolda},\ and\ \citenamefont {March-Russell}}]{Babu:1997st}%
  \BibitemOpen
  \bibfield  {author} {\bibinfo {author} {\bibfnamefont {K.~S.}\ \bibnamefont
  {Babu}}, \bibinfo {author} {\bibfnamefont {C.~F.}\ \bibnamefont {Kolda}}, \
  and\ \bibinfo {author} {\bibfnamefont {J.}~\bibnamefont {March-Russell}},\
  }\href {\doibase 10.1103/PhysRevD.57.6788} {\bibfield  {journal} {\bibinfo
  {journal} {Phys. Rev.}\ }\textbf {\bibinfo {volume} {D57}},\ \bibinfo {pages}
  {6788} (\bibinfo {year} {1998})},\ \Eprint
  {http://arxiv.org/abs/hep-ph/9710441} {arXiv:hep-ph/9710441 [hep-ph]}
  \BibitemShut {NoStop}%
%%CITATION = HEP-PH/9710441;%%
\bibitem [{\citenamefont {Chun}\ \emph {et~al.}(2011)\citenamefont {Chun},
  \citenamefont {Park},\ and\ \citenamefont {Scopel}}]{Chun:2010ve}%
  \BibitemOpen
  \bibfield  {author} {\bibinfo {author} {\bibfnamefont {E.~J.}\ \bibnamefont
  {Chun}}, \bibinfo {author} {\bibfnamefont {J.-C.}\ \bibnamefont {Park}}, \
  and\ \bibinfo {author} {\bibfnamefont {S.}~\bibnamefont {Scopel}},\ }\href
  {\doibase 10.1007/JHEP02(2011)100} {\bibfield  {journal} {\bibinfo  {journal}
  {JHEP}\ }\textbf {\bibinfo {volume} {02}},\ \bibinfo {pages} {100} (\bibinfo
  {year} {2011})},\ \Eprint {http://arxiv.org/abs/1011.3300} {arXiv:1011.3300
  [hep-ph]} \BibitemShut {NoStop}%
%%CITATION = ARXIV:1011.3300;%%
\bibitem [{\citenamefont {Patrignani}\ \emph {et~al.}(2016)\citenamefont
  {Patrignani} \emph {et~al.}}]{Patrignani:2016xqp}%
  \BibitemOpen
  \bibfield  {author} {\bibinfo {author} {\bibfnamefont {C.}~\bibnamefont
  {Patrignani}} \emph {et~al.} (\bibinfo {collaboration} {Particle Data
  Group}),\ }\href {\doibase 10.1088/1674-1137/40/10/100001} {\bibfield
  {journal} {\bibinfo  {journal} {Chin. Phys.}\ }\textbf {\bibinfo {volume}
  {C40}},\ \bibinfo {pages} {100001} (\bibinfo {year} {2016})}\BibitemShut
  {NoStop}%
%%CITATION = CHPHD,C40,100001;%%
\bibitem [{\citenamefont {Kumar}\ and\ \citenamefont
  {Wells}(2006)}]{Kumar:2006gm}%
  \BibitemOpen
  \bibfield  {author} {\bibinfo {author} {\bibfnamefont {J.}~\bibnamefont
  {Kumar}}\ and\ \bibinfo {author} {\bibfnamefont {J.~D.}\ \bibnamefont
  {Wells}},\ }\href {\doibase 10.1103/PhysRevD.74.115017} {\bibfield  {journal}
  {\bibinfo  {journal} {Phys. Rev.}\ }\textbf {\bibinfo {volume} {D74}},\
  \bibinfo {pages} {115017} (\bibinfo {year} {2006})},\ \Eprint
  {http://arxiv.org/abs/hep-ph/0606183} {arXiv:hep-ph/0606183 [hep-ph]}
  \BibitemShut {NoStop}%
%%CITATION = HEP-PH/0606183;%%
\bibitem [{\citenamefont {Ge}\ \emph {et~al.}(2017)\citenamefont {Ge},
  \citenamefont {Lindner},\ and\ \citenamefont {Rodejohann}}]{Ge:2017poy}%
  \BibitemOpen
  \bibfield  {author} {\bibinfo {author} {\bibfnamefont {S.-F.}\ \bibnamefont
  {Ge}}, \bibinfo {author} {\bibfnamefont {M.}~\bibnamefont {Lindner}}, \ and\
  \bibinfo {author} {\bibfnamefont {W.}~\bibnamefont {Rodejohann}},\ }\href
  {\doibase 10.1016/j.physletb.2017.06.020} {\bibfield  {journal} {\bibinfo
  {journal} {Phys. Lett.}\ }\textbf {\bibinfo {volume} {B772}},\ \bibinfo
  {pages} {164} (\bibinfo {year} {2017})},\ \Eprint
  {http://arxiv.org/abs/1702.02617} {arXiv:1702.02617 [hep-ph]} \BibitemShut
  {NoStop}%
%%CITATION = ARXIV:1702.02617;%%
\bibitem [{\citenamefont {Altmannshofer}\ \emph
  {et~al.}(2014{\natexlab{b}})\citenamefont {Altmannshofer}, \citenamefont
  {Gori}, \citenamefont {Pospelov},\ and\ \citenamefont
  {Yavin}}]{Altmannshofer:2014pba}%
  \BibitemOpen
  \bibfield  {author} {\bibinfo {author} {\bibfnamefont {W.}~\bibnamefont
  {Altmannshofer}}, \bibinfo {author} {\bibfnamefont {S.}~\bibnamefont {Gori}},
  \bibinfo {author} {\bibfnamefont {M.}~\bibnamefont {Pospelov}}, \ and\
  \bibinfo {author} {\bibfnamefont {I.}~\bibnamefont {Yavin}},\ }\href
  {\doibase 10.1103/PhysRevLett.113.091801} {\bibfield  {journal} {\bibinfo
  {journal} {Phys. Rev. Lett.}\ }\textbf {\bibinfo {volume} {113}},\ \bibinfo
  {pages} {091801} (\bibinfo {year} {2014}{\natexlab{b}})},\ \Eprint
  {http://arxiv.org/abs/1406.2332} {arXiv:1406.2332 [hep-ph]} \BibitemShut
  {NoStop}%
%%CITATION = ARXIV:1406.2332;%%
\bibitem [{\citenamefont {Blake}\ \emph {et~al.}(2017)\citenamefont {Blake},
  \citenamefont {Lanfranchi},\ and\ \citenamefont {Straub}}]{Blake:2016olu}%
  \BibitemOpen
  \bibfield  {author} {\bibinfo {author} {\bibfnamefont {T.}~\bibnamefont
  {Blake}}, \bibinfo {author} {\bibfnamefont {G.}~\bibnamefont {Lanfranchi}}, \
  and\ \bibinfo {author} {\bibfnamefont {D.~M.}\ \bibnamefont {Straub}},\
  }\href {\doibase 10.1016/j.ppnp.2016.10.001} {\bibfield  {journal} {\bibinfo
  {journal} {Prog. Part. Nucl. Phys.}\ }\textbf {\bibinfo {volume} {92}},\
  \bibinfo {pages} {50} (\bibinfo {year} {2017})},\ \Eprint
  {http://arxiv.org/abs/1606.00916} {arXiv:1606.00916 [hep-ph]} \BibitemShut
  {NoStop}%
%%CITATION = ARXIV:1606.00916;%%
\bibitem [{\citenamefont {Charles}\ \emph {et~al.}(2015)\citenamefont {Charles}
  \emph {et~al.}}]{Charles:2015gya}%
  \BibitemOpen
  \bibfield  {author} {\bibinfo {author} {\bibfnamefont {J.}~\bibnamefont
  {Charles}} \emph {et~al.},\ }\href {\doibase 10.1103/PhysRevD.91.073007}
  {\bibfield  {journal} {\bibinfo  {journal} {Phys. Rev.}\ }\textbf {\bibinfo
  {volume} {D91}},\ \bibinfo {pages} {073007} (\bibinfo {year} {2015})},\
  \Eprint {http://arxiv.org/abs/1501.05013} {arXiv:1501.05013 [hep-ph]}
  \BibitemShut {NoStop}%
%%CITATION = ARXIV:1501.05013;%%
\bibitem [{\citenamefont {Bazavov}\ \emph {et~al.}(2016)\citenamefont {Bazavov}
  \emph {et~al.}}]{Bazavov:2016nty}%
  \BibitemOpen
  \bibfield  {author} {\bibinfo {author} {\bibfnamefont {A.}~\bibnamefont
  {Bazavov}} \emph {et~al.} (\bibinfo {collaboration} {Fermilab Lattice,
  MILC}),\ }\href {\doibase 10.1103/PhysRevD.93.113016} {\bibfield  {journal}
  {\bibinfo  {journal} {Phys. Rev.}\ }\textbf {\bibinfo {volume} {D93}},\
  \bibinfo {pages} {113016} (\bibinfo {year} {2016})},\ \Eprint
  {http://arxiv.org/abs/1602.03560} {arXiv:1602.03560 [hep-lat]} \BibitemShut
  {NoStop}%
%%CITATION = ARXIV:1602.03560;%%
\bibitem [{\citenamefont {Khachatryan}\ \emph {et~al.}(2016)\citenamefont
  {Khachatryan} \emph {et~al.}}]{Khachatryan:2015dcf}%
  \BibitemOpen
  \bibfield  {author} {\bibinfo {author} {\bibfnamefont {V.}~\bibnamefont
  {Khachatryan}} \emph {et~al.} (\bibinfo {collaboration} {CMS}),\ }\href
  {\doibase 10.1103/PhysRevLett.116.071801} {\bibfield  {journal} {\bibinfo
  {journal} {Phys. Rev. Lett.}\ }\textbf {\bibinfo {volume} {116}},\ \bibinfo
  {pages} {071801} (\bibinfo {year} {2016})},\ \Eprint
  {http://arxiv.org/abs/1512.01224} {arXiv:1512.01224 [hep-ex]} \BibitemShut
  {NoStop}%
%%CITATION = ARXIV:1512.01224;%%
\bibitem [{\citenamefont {Aaboud}\ \emph {et~al.}(2017)\citenamefont {Aaboud}
  \emph {et~al.}}]{Aaboud:2017buh}%
  \BibitemOpen
  \bibfield  {author} {\bibinfo {author} {\bibfnamefont {M.}~\bibnamefont
  {Aaboud}} \emph {et~al.} (\bibinfo {collaboration} {ATLAS}),\ }\href
  {\doibase 10.1007/JHEP10(2017)182} {\bibfield  {journal} {\bibinfo  {journal}
  {JHEP}\ }\textbf {\bibinfo {volume} {10}},\ \bibinfo {pages} {182} (\bibinfo
  {year} {2017})},\ \Eprint {http://arxiv.org/abs/1707.02424} {arXiv:1707.02424
  [hep-ex]} \BibitemShut {NoStop}%
%%CITATION = ARXIV:1707.02424;%%
\bibitem [{\citenamefont {Arcadi}\ \emph {et~al.}(2017)\citenamefont {Arcadi},
  \citenamefont {Dutra}, \citenamefont {Ghosh}, \citenamefont {Lindner},
  \citenamefont {Mambrini}, \citenamefont {Pierre}, \citenamefont {Profumo},\
  and\ \citenamefont {Queiroz}}]{Arcadi:2017kky}%
  \BibitemOpen
  \bibfield  {author} {\bibinfo {author} {\bibfnamefont {G.}~\bibnamefont
  {Arcadi}}, \bibinfo {author} {\bibfnamefont {M.}~\bibnamefont {Dutra}},
  \bibinfo {author} {\bibfnamefont {P.}~\bibnamefont {Ghosh}}, \bibinfo
  {author} {\bibfnamefont {M.}~\bibnamefont {Lindner}}, \bibinfo {author}
  {\bibfnamefont {Y.}~\bibnamefont {Mambrini}}, \bibinfo {author}
  {\bibfnamefont {M.}~\bibnamefont {Pierre}}, \bibinfo {author} {\bibfnamefont
  {S.}~\bibnamefont {Profumo}}, \ and\ \bibinfo {author} {\bibfnamefont
  {F.~S.}\ \bibnamefont {Queiroz}},\ }\href@noop {} {\  (\bibinfo {year}
  {2017})},\ \Eprint {http://arxiv.org/abs/1703.07364} {arXiv:1703.07364
  [hep-ph]} \BibitemShut {NoStop}%
%%CITATION = ARXIV:1703.07364;%%
\bibitem [{\citenamefont {Griest}\ and\ \citenamefont
  {Seckel}(1991)}]{Griest:1990kh}%
  \BibitemOpen
  \bibfield  {author} {\bibinfo {author} {\bibfnamefont {K.}~\bibnamefont
  {Griest}}\ and\ \bibinfo {author} {\bibfnamefont {D.}~\bibnamefont
  {Seckel}},\ }\href {\doibase 10.1103/PhysRevD.43.3191} {\bibfield  {journal}
  {\bibinfo  {journal} {Phys. Rev.}\ }\textbf {\bibinfo {volume} {D43}},\
  \bibinfo {pages} {3191} (\bibinfo {year} {1991})}\BibitemShut {NoStop}%
%%CITATION = PHRVA,D43,3191;%%
\bibitem [{\citenamefont {Belanger}\ \emph {et~al.}(2014)\citenamefont
  {Belanger}, \citenamefont {Boudjema}, \citenamefont {Pukhov},\ and\
  \citenamefont {Semenov}}]{Belanger:2013oya}%
  \BibitemOpen
  \bibfield  {author} {\bibinfo {author} {\bibfnamefont {G.}~\bibnamefont
  {Belanger}}, \bibinfo {author} {\bibfnamefont {F.}~\bibnamefont {Boudjema}},
  \bibinfo {author} {\bibfnamefont {A.}~\bibnamefont {Pukhov}}, \ and\ \bibinfo
  {author} {\bibfnamefont {A.}~\bibnamefont {Semenov}},\ }\href {\doibase
  10.1016/j.cpc.2013.10.016} {\bibfield  {journal} {\bibinfo  {journal}
  {Comput. Phys. Commun.}\ }\textbf {\bibinfo {volume} {185}},\ \bibinfo
  {pages} {960} (\bibinfo {year} {2014})},\ \Eprint
  {http://arxiv.org/abs/1305.0237} {arXiv:1305.0237 [hep-ph]} \BibitemShut
  {NoStop}%
%%CITATION = ARXIV:1305.0237;%%
\bibitem [{\citenamefont {Dutta}\ \emph {et~al.}(2017)\citenamefont {Dutta},
  \citenamefont {Jimenez},\ and\ \citenamefont {Zavala}}]{Dutta:2016htz}%
  \BibitemOpen
  \bibfield  {author} {\bibinfo {author} {\bibfnamefont {B.}~\bibnamefont
  {Dutta}}, \bibinfo {author} {\bibfnamefont {E.}~\bibnamefont {Jimenez}}, \
  and\ \bibinfo {author} {\bibfnamefont {I.}~\bibnamefont {Zavala}},\ }\href
  {\doibase 10.1088/1475-7516/2017/06/032} {\bibfield  {journal} {\bibinfo
  {journal} {JCAP}\ }\textbf {\bibinfo {volume} {1706}},\ \bibinfo {pages}
  {032} (\bibinfo {year} {2017})},\ \Eprint {http://arxiv.org/abs/1612.05553}
  {arXiv:1612.05553 [hep-ph]} \BibitemShut {NoStop}%
%%CITATION = ARXIV:1612.05553;%%
\bibitem [{\citenamefont {Dutra}\ \emph {et~al.}(2018)\citenamefont {Dutra},
  \citenamefont {Lindner}, \citenamefont {Profumo}, \citenamefont {Queiroz},
  \citenamefont {Rodejohann},\ and\ \citenamefont {Siqueira}}]{Dutra:2018gmv}%
  \BibitemOpen
  \bibfield  {author} {\bibinfo {author} {\bibfnamefont {M.}~\bibnamefont
  {Dutra}}, \bibinfo {author} {\bibfnamefont {M.}~\bibnamefont {Lindner}},
  \bibinfo {author} {\bibfnamefont {S.}~\bibnamefont {Profumo}}, \bibinfo
  {author} {\bibfnamefont {F.~S.}\ \bibnamefont {Queiroz}}, \bibinfo {author}
  {\bibfnamefont {W.}~\bibnamefont {Rodejohann}}, \ and\ \bibinfo {author}
  {\bibfnamefont {C.}~\bibnamefont {Siqueira}},\ }\href@noop {} {\  (\bibinfo
  {year} {2018})},\ \Eprint {http://arxiv.org/abs/1801.05447} {arXiv:1801.05447
  [hep-ph]} \BibitemShut {NoStop}%
%%CITATION = ARXIV:1801.05447;%%
\bibitem [{\citenamefont {Angle}\ \emph {et~al.}(2011)\citenamefont {Angle}
  \emph {et~al.}}]{Angle:2011th}%
  \BibitemOpen
  \bibfield  {author} {\bibinfo {author} {\bibfnamefont {J.}~\bibnamefont
  {Angle}} \emph {et~al.} (\bibinfo {collaboration} {XENON10}),\ }\href
  {\doibase 10.1103/PhysRevLett.110.249901, 10.1103/PhysRevLett.107.051301}
  {\bibfield  {journal} {\bibinfo  {journal} {Phys. Rev. Lett.}\ }\textbf
  {\bibinfo {volume} {107}},\ \bibinfo {pages} {051301} (\bibinfo {year}
  {2011})},\ \bibinfo {note} {[Erratum: Phys. Rev. Lett.110,249901(2013)]},\
  \Eprint {http://arxiv.org/abs/1104.3088} {arXiv:1104.3088 [astro-ph.CO]}
  \BibitemShut {NoStop}%
%%CITATION = ARXIV:1104.3088;%%
\bibitem [{\citenamefont {Akerib}\ \emph {et~al.}(2016)\citenamefont {Akerib}
  \emph {et~al.}}]{Akerib:2015rjg}%
  \BibitemOpen
  \bibfield  {author} {\bibinfo {author} {\bibfnamefont {D.~S.}\ \bibnamefont
  {Akerib}} \emph {et~al.} (\bibinfo {collaboration} {LUX}),\ }\href {\doibase
  10.1103/PhysRevLett.116.161301} {\bibfield  {journal} {\bibinfo  {journal}
  {Phys. Rev. Lett.}\ }\textbf {\bibinfo {volume} {116}},\ \bibinfo {pages}
  {161301} (\bibinfo {year} {2016})},\ \Eprint
  {http://arxiv.org/abs/1512.03506} {arXiv:1512.03506 [astro-ph.CO]}
  \BibitemShut {NoStop}%
%%CITATION = ARXIV:1512.03506;%%
\bibitem [{\citenamefont {Aguilar-Arevalo}\ \emph {et~al.}(2017)\citenamefont
  {Aguilar-Arevalo} \emph {et~al.}}]{Aguilar-Arevalo:2016zop}%
  \BibitemOpen
  \bibfield  {author} {\bibinfo {author} {\bibfnamefont {A.}~\bibnamefont
  {Aguilar-Arevalo}} \emph {et~al.} (\bibinfo {collaboration} {DAMIC}),\ }\href
  {\doibase 10.1103/PhysRevLett.118.141803} {\bibfield  {journal} {\bibinfo
  {journal} {Phys. Rev. Lett.}\ }\textbf {\bibinfo {volume} {118}},\ \bibinfo
  {pages} {141803} (\bibinfo {year} {2017})},\ \Eprint
  {http://arxiv.org/abs/1611.03066} {arXiv:1611.03066 [astro-ph.CO]}
  \BibitemShut {NoStop}%
%%CITATION = ARXIV:1611.03066;%%
\bibitem [{\citenamefont {Aguilar-Arevalo}\ \emph {et~al.}(2016)\citenamefont
  {Aguilar-Arevalo} \emph {et~al.}}]{Aguilar-Arevalo:2016ndq}%
  \BibitemOpen
  \bibfield  {author} {\bibinfo {author} {\bibfnamefont {A.}~\bibnamefont
  {Aguilar-Arevalo}} \emph {et~al.} (\bibinfo {collaboration} {DAMIC}),\ }\href
  {\doibase 10.1103/PhysRevD.94.082006} {\bibfield  {journal} {\bibinfo
  {journal} {Phys. Rev.}\ }\textbf {\bibinfo {volume} {D94}},\ \bibinfo {pages}
  {082006} (\bibinfo {year} {2016})},\ \Eprint
  {http://arxiv.org/abs/1607.07410} {arXiv:1607.07410 [astro-ph.CO]}
  \BibitemShut {NoStop}%
%%CITATION = ARXIV:1607.07410;%%
\bibitem [{\citenamefont {Tan}\ \emph {et~al.}(2016)\citenamefont {Tan} \emph
  {et~al.}}]{Tan:2016zwf}%
  \BibitemOpen
  \bibfield  {author} {\bibinfo {author} {\bibfnamefont {A.}~\bibnamefont
  {Tan}} \emph {et~al.} (\bibinfo {collaboration} {PandaX-II}),\ }\href
  {\doibase 10.1103/PhysRevLett.117.121303} {\bibfield  {journal} {\bibinfo
  {journal} {Phys. Rev. Lett.}\ }\textbf {\bibinfo {volume} {117}},\ \bibinfo
  {pages} {121303} (\bibinfo {year} {2016})},\ \Eprint
  {http://arxiv.org/abs/1607.07400} {arXiv:1607.07400 [hep-ex]} \BibitemShut
  {NoStop}%
%%CITATION = ARXIV:1607.07400;%%
\bibitem [{\citenamefont {Agnese}\ \emph {et~al.}(2018)\citenamefont {Agnese}
  \emph {et~al.}}]{Agnese:2017njq}%
  \BibitemOpen
  \bibfield  {author} {\bibinfo {author} {\bibfnamefont {R.}~\bibnamefont
  {Agnese}} \emph {et~al.} (\bibinfo {collaboration} {SuperCDMS}),\ }\href
  {\doibase 10.1103/PhysRevLett.120.061802} {\bibfield  {journal} {\bibinfo
  {journal} {Phys. Rev. Lett.}\ }\textbf {\bibinfo {volume} {120}},\ \bibinfo
  {pages} {061802} (\bibinfo {year} {2018})},\ \Eprint
  {http://arxiv.org/abs/1708.08869} {arXiv:1708.08869 [hep-ex]} \BibitemShut
  {NoStop}%
%%CITATION = ARXIV:1708.08869;%%
\bibitem [{\citenamefont {Amole}\ \emph {et~al.}(2017)\citenamefont {Amole}
  \emph {et~al.}}]{Amole:2017dex}%
  \BibitemOpen
  \bibfield  {author} {\bibinfo {author} {\bibfnamefont {C.}~\bibnamefont
  {Amole}} \emph {et~al.} (\bibinfo {collaboration} {PICO}),\ }\href {\doibase
  10.1103/PhysRevLett.118.251301} {\bibfield  {journal} {\bibinfo  {journal}
  {Phys. Rev. Lett.}\ }\textbf {\bibinfo {volume} {118}},\ \bibinfo {pages}
  {251301} (\bibinfo {year} {2017})},\ \Eprint
  {http://arxiv.org/abs/1702.07666} {arXiv:1702.07666 [astro-ph.CO]}
  \BibitemShut {NoStop}%
%%CITATION = ARXIV:1702.07666;%%
\bibitem [{\citenamefont {Aprile}\ \emph
  {et~al.}(2017{\natexlab{a}})\citenamefont {Aprile} \emph
  {et~al.}}]{Aprile:2017iyp}%
  \BibitemOpen
  \bibfield  {author} {\bibinfo {author} {\bibfnamefont {E.}~\bibnamefont
  {Aprile}} \emph {et~al.} (\bibinfo {collaboration} {XENON}),\ }\href
  {\doibase 10.1103/PhysRevLett.119.181301} {\bibfield  {journal} {\bibinfo
  {journal} {Phys. Rev. Lett.}\ }\textbf {\bibinfo {volume} {119}},\ \bibinfo
  {pages} {181301} (\bibinfo {year} {2017}{\natexlab{a}})},\ \Eprint
  {http://arxiv.org/abs/1705.06655} {arXiv:1705.06655 [astro-ph.CO]}
  \BibitemShut {NoStop}%
%%CITATION = ARXIV:1705.06655;%%
\bibitem [{\citenamefont {Aprile}\ \emph
  {et~al.}(2017{\natexlab{b}})\citenamefont {Aprile} \emph
  {et~al.}}]{Aprile:2017ngb}%
  \BibitemOpen
  \bibfield  {author} {\bibinfo {author} {\bibfnamefont {E.}~\bibnamefont
  {Aprile}} \emph {et~al.} (\bibinfo {collaboration} {XENON}),\ }\href
  {\doibase 10.1103/PhysRevD.96.022008} {\bibfield  {journal} {\bibinfo
  {journal} {Phys. Rev.}\ }\textbf {\bibinfo {volume} {D96}},\ \bibinfo {pages}
  {022008} (\bibinfo {year} {2017}{\natexlab{b}})},\ \Eprint
  {http://arxiv.org/abs/1705.05830} {arXiv:1705.05830 [hep-ex]} \BibitemShut
  {NoStop}%
%%CITATION = ARXIV:1705.05830;%%
\bibitem [{\citenamefont {Aprile}\ \emph
  {et~al.}(2017{\natexlab{c}})\citenamefont {Aprile} \emph
  {et~al.}}]{Aprile:2017yea}%
  \BibitemOpen
  \bibfield  {author} {\bibinfo {author} {\bibfnamefont {E.}~\bibnamefont
  {Aprile}} \emph {et~al.} (\bibinfo {collaboration} {XENON}),\ }\href
  {\doibase 10.1103/PhysRevLett.118.101101} {\bibfield  {journal} {\bibinfo
  {journal} {Phys. Rev. Lett.}\ }\textbf {\bibinfo {volume} {118}},\ \bibinfo
  {pages} {101101} (\bibinfo {year} {2017}{\natexlab{c}})},\ \Eprint
  {http://arxiv.org/abs/1701.00769} {arXiv:1701.00769 [astro-ph.CO]}
  \BibitemShut {NoStop}%
%%CITATION = ARXIV:1701.00769;%%
\bibitem [{\citenamefont {Cui}\ \emph {et~al.}(2017)\citenamefont {Cui} \emph
  {et~al.}}]{Cui:2017nnn}%
  \BibitemOpen
  \bibfield  {author} {\bibinfo {author} {\bibfnamefont {X.}~\bibnamefont
  {Cui}} \emph {et~al.} (\bibinfo {collaboration} {PandaX-II}),\ }\href
  {\doibase 10.1103/PhysRevLett.119.181302} {\bibfield  {journal} {\bibinfo
  {journal} {Phys. Rev. Lett.}\ }\textbf {\bibinfo {volume} {119}},\ \bibinfo
  {pages} {181302} (\bibinfo {year} {2017})},\ \Eprint
  {http://arxiv.org/abs/1708.06917} {arXiv:1708.06917 [astro-ph.CO]}
  \BibitemShut {NoStop}%
%%CITATION = ARXIV:1708.06917;%%
\bibitem [{\citenamefont {Akerib}\ \emph {et~al.}(2017)\citenamefont {Akerib}
  \emph {et~al.}}]{Akerib:2017kat}%
  \BibitemOpen
  \bibfield  {author} {\bibinfo {author} {\bibfnamefont {D.~S.}\ \bibnamefont
  {Akerib}} \emph {et~al.} (\bibinfo {collaboration} {LUX}),\ }\href {\doibase
  10.1103/PhysRevLett.118.251302} {\bibfield  {journal} {\bibinfo  {journal}
  {Phys. Rev. Lett.}\ }\textbf {\bibinfo {volume} {118}},\ \bibinfo {pages}
  {251302} (\bibinfo {year} {2017})},\ \Eprint
  {http://arxiv.org/abs/1705.03380} {arXiv:1705.03380 [astro-ph.CO]}
  \BibitemShut {NoStop}%
%%CITATION = ARXIV:1705.03380;%%
\bibitem [{\citenamefont {Agnes}\ \emph
  {et~al.}(2018{\natexlab{a}})\citenamefont {Agnes} \emph
  {et~al.}}]{Agnes:2018oej}%
  \BibitemOpen
  \bibfield  {author} {\bibinfo {author} {\bibfnamefont {P.}~\bibnamefont
  {Agnes}} \emph {et~al.} (\bibinfo {collaboration} {DarkSide}),\ }\href@noop
  {} {\  (\bibinfo {year} {2018}{\natexlab{a}})},\ \Eprint
  {http://arxiv.org/abs/1802.06998} {arXiv:1802.06998 [astro-ph.CO]}
  \BibitemShut {NoStop}%
%%CITATION = ARXIV:1802.06998;%%
\bibitem [{\citenamefont {Agnes}\ \emph
  {et~al.}(2018{\natexlab{b}})\citenamefont {Agnes} \emph
  {et~al.}}]{Agnes:2018ves}%
  \BibitemOpen
  \bibfield  {author} {\bibinfo {author} {\bibfnamefont {P.}~\bibnamefont
  {Agnes}} \emph {et~al.} (\bibinfo {collaboration} {DarkSide}),\ }\href@noop
  {} {\  (\bibinfo {year} {2018}{\natexlab{b}})},\ \Eprint
  {http://arxiv.org/abs/1802.06994} {arXiv:1802.06994 [astro-ph.HE]}
  \BibitemShut {NoStop}%
%%CITATION = ARXIV:1802.06994;%%
\bibitem [{\citenamefont {Feng}\ \emph {et~al.}(2013)\citenamefont {Feng},
  \citenamefont {Kumar},\ and\ \citenamefont {Sanford}}]{Feng:2013fyw}%
  \BibitemOpen
  \bibfield  {author} {\bibinfo {author} {\bibfnamefont {J.~L.}\ \bibnamefont
  {Feng}}, \bibinfo {author} {\bibfnamefont {J.}~\bibnamefont {Kumar}}, \ and\
  \bibinfo {author} {\bibfnamefont {D.}~\bibnamefont {Sanford}},\ }\href
  {\doibase 10.1103/PhysRevD.88.015021} {\bibfield  {journal} {\bibinfo
  {journal} {Phys. Rev.}\ }\textbf {\bibinfo {volume} {D88}},\ \bibinfo {pages}
  {015021} (\bibinfo {year} {2013})},\ \Eprint {http://arxiv.org/abs/1306.2315}
  {arXiv:1306.2315 [hep-ph]} \BibitemShut {NoStop}%
%%CITATION = ARXIV:1306.2315;%%
\bibitem [{\citenamefont {Aprile}\ \emph {et~al.}(2016)\citenamefont {Aprile}
  \emph {et~al.}}]{Aprile:2015uzo}%
  \BibitemOpen
  \bibfield  {author} {\bibinfo {author} {\bibfnamefont {E.}~\bibnamefont
  {Aprile}} \emph {et~al.} (\bibinfo {collaboration} {XENON}),\ }\href
  {\doibase 10.1088/1475-7516/2016/04/027} {\bibfield  {journal} {\bibinfo
  {journal} {JCAP}\ }\textbf {\bibinfo {volume} {1604}},\ \bibinfo {pages}
  {027} (\bibinfo {year} {2016})},\ \Eprint {http://arxiv.org/abs/1512.07501}
  {arXiv:1512.07501 [physics.ins-det]} \BibitemShut {NoStop}%
%%CITATION = ARXIV:1512.07501;%%
\bibitem [{\citenamefont {Aalbers}\ \emph {et~al.}(2016)\citenamefont {Aalbers}
  \emph {et~al.}}]{Aalbers:2016jon}%
  \BibitemOpen
  \bibfield  {author} {\bibinfo {author} {\bibfnamefont {J.}~\bibnamefont
  {Aalbers}} \emph {et~al.} (\bibinfo {collaboration} {DARWIN}),\ }\href
  {\doibase 10.1088/1475-7516/2016/11/017} {\bibfield  {journal} {\bibinfo
  {journal} {JCAP}\ }\textbf {\bibinfo {volume} {1611}},\ \bibinfo {pages}
  {017} (\bibinfo {year} {2016})},\ \Eprint {http://arxiv.org/abs/1606.07001}
  {arXiv:1606.07001 [astro-ph.IM]} \BibitemShut {NoStop}%
%%CITATION = ARXIV:1606.07001;%%
\bibitem [{\citenamefont {Hooper}\ \emph {et~al.}(2013)\citenamefont {Hooper},
  \citenamefont {Kelso},\ and\ \citenamefont {Queiroz}}]{Hooper:2012sr}%
  \BibitemOpen
  \bibfield  {author} {\bibinfo {author} {\bibfnamefont {D.}~\bibnamefont
  {Hooper}}, \bibinfo {author} {\bibfnamefont {C.}~\bibnamefont {Kelso}}, \
  and\ \bibinfo {author} {\bibfnamefont {F.~S.}\ \bibnamefont {Queiroz}},\
  }\href {\doibase 10.1016/j.astropartphys.2013.04.007} {\bibfield  {journal}
  {\bibinfo  {journal} {Astropart. Phys.}\ }\textbf {\bibinfo {volume} {46}},\
  \bibinfo {pages} {55} (\bibinfo {year} {2013})},\ \Eprint
  {http://arxiv.org/abs/1209.3015} {arXiv:1209.3015 [astro-ph.HE]} \BibitemShut
  {NoStop}%
%%CITATION = ARXIV:1209.3015;%%
\bibitem [{\citenamefont {Tavani}\ \emph {et~al.}(2017)\citenamefont {Tavani}
  \emph {et~al.}}]{DeAngelis:2017gra}%
  \BibitemOpen
  \bibfield  {author} {\bibinfo {author} {\bibfnamefont {M.}~\bibnamefont
  {Tavani}} \emph {et~al.} (\bibinfo {collaboration} {e-ASTROGAM}),\
  }\href@noop {} {\  (\bibinfo {year} {2017})},\ \Eprint
  {http://arxiv.org/abs/1711.01265} {arXiv:1711.01265 [astro-ph.HE]}
  \BibitemShut {NoStop}%
%%CITATION = ARXIV:1711.01265;%%
\bibitem [{\citenamefont {Acharya}\ \emph {et~al.}(2017)\citenamefont {Acharya}
  \emph {et~al.}}]{Acharya:2017ttl}%
  \BibitemOpen
  \bibfield  {author} {\bibinfo {author} {\bibfnamefont {B.~S.}\ \bibnamefont
  {Acharya}} \emph {et~al.} (\bibinfo {collaboration} {Cherenkov Telescope
  Array Consortium}),\ }\href@noop {} {\  (\bibinfo {year} {2017})},\ \Eprint
  {http://arxiv.org/abs/1709.07997} {arXiv:1709.07997 [astro-ph.IM]}
  \BibitemShut {NoStop}%
%%CITATION = ARXIV:1709.07997;%%
\bibitem [{\citenamefont {Li}\ \emph {et~al.}(2018)\citenamefont {Li} \emph
  {et~al.}}]{Li:2018kgy}%
  \BibitemOpen
  \bibfield  {author} {\bibinfo {author} {\bibfnamefont {S.}~\bibnamefont {Li}}
  \emph {et~al.},\ }\href@noop {} {\  (\bibinfo {year} {2018})},\ \Eprint
  {http://arxiv.org/abs/1805.06612} {arXiv:1805.06612 [astro-ph.HE]}
  \BibitemShut {NoStop}%
%%CITATION = ARXIV:1805.06612;%%
\bibitem [{\citenamefont {Bringmann}\ \emph {et~al.}(2014)\citenamefont
  {Bringmann}, \citenamefont {Vollmann},\ and\ \citenamefont
  {Weniger}}]{Bringmann:2014lpa}%
  \BibitemOpen
  \bibfield  {author} {\bibinfo {author} {\bibfnamefont {T.}~\bibnamefont
  {Bringmann}}, \bibinfo {author} {\bibfnamefont {M.}~\bibnamefont {Vollmann}},
  \ and\ \bibinfo {author} {\bibfnamefont {C.}~\bibnamefont {Weniger}},\ }\href
  {\doibase 10.1103/PhysRevD.90.123001} {\bibfield  {journal} {\bibinfo
  {journal} {Phys. Rev.}\ }\textbf {\bibinfo {volume} {D90}},\ \bibinfo {pages}
  {123001} (\bibinfo {year} {2014})},\ \Eprint {http://arxiv.org/abs/1406.6027}
  {arXiv:1406.6027 [astro-ph.HE]} \BibitemShut {NoStop}%
%%CITATION = ARXIV:1406.6027;%%
\bibitem [{\citenamefont {Giesen}\ \emph {et~al.}(2015)\citenamefont {Giesen},
  \citenamefont {Boudaud}, \citenamefont {Génolini}, \citenamefont {Poulin},
  \citenamefont {Cirelli}, \citenamefont {Salati},\ and\ \citenamefont
  {Serpico}}]{Giesen:2015ufa}%
  \BibitemOpen
  \bibfield  {author} {\bibinfo {author} {\bibfnamefont {G.}~\bibnamefont
  {Giesen}}, \bibinfo {author} {\bibfnamefont {M.}~\bibnamefont {Boudaud}},
  \bibinfo {author} {\bibfnamefont {Y.}~\bibnamefont {Génolini}}, \bibinfo
  {author} {\bibfnamefont {V.}~\bibnamefont {Poulin}}, \bibinfo {author}
  {\bibfnamefont {M.}~\bibnamefont {Cirelli}}, \bibinfo {author} {\bibfnamefont
  {P.}~\bibnamefont {Salati}}, \ and\ \bibinfo {author} {\bibfnamefont {P.~D.}\
  \bibnamefont {Serpico}},\ }\href {\doibase 10.1088/1475-7516/2015/09/023,
  10.1088/1475-7516/2015/9/023} {\bibfield  {journal} {\bibinfo  {journal}
  {JCAP}\ }\textbf {\bibinfo {volume} {1509}},\ \bibinfo {pages} {023}
  (\bibinfo {year} {2015})},\ \Eprint {http://arxiv.org/abs/1504.04276}
  {arXiv:1504.04276 [astro-ph.HE]} \BibitemShut {NoStop}%
%%CITATION = ARXIV:1504.04276;%%
\bibitem [{\citenamefont {Profumo}\ \emph {et~al.}(2018)\citenamefont
  {Profumo}, \citenamefont {Queiroz}, \citenamefont {Silk},\ and\ \citenamefont
  {Siqueira}}]{Profumo:2017obk}%
  \BibitemOpen
  \bibfield  {author} {\bibinfo {author} {\bibfnamefont {S.}~\bibnamefont
  {Profumo}}, \bibinfo {author} {\bibfnamefont {F.~S.}\ \bibnamefont
  {Queiroz}}, \bibinfo {author} {\bibfnamefont {J.}~\bibnamefont {Silk}}, \
  and\ \bibinfo {author} {\bibfnamefont {C.}~\bibnamefont {Siqueira}},\ }\href
  {\doibase 10.1088/1475-7516/2018/03/010} {\bibfield  {journal} {\bibinfo
  {journal} {JCAP}\ }\textbf {\bibinfo {volume} {1803}},\ \bibinfo {pages}
  {010} (\bibinfo {year} {2018})},\ \Eprint {http://arxiv.org/abs/1711.03133}
  {arXiv:1711.03133 [hep-ph]} \BibitemShut {NoStop}%
%%CITATION = ARXIV:1711.03133;%%
\bibitem [{\citenamefont {Balázs}\ \emph {et~al.}(2017)\citenamefont
  {Balázs}, \citenamefont {Conrad}, \citenamefont {Farmer}, \citenamefont
  {Jacques}, \citenamefont {Li}, \citenamefont {Meyer}, \citenamefont
  {Queiroz},\ and\ \citenamefont {Sánchez-Conde}}]{Balazs:2017hxh}%
  \BibitemOpen
  \bibfield  {author} {\bibinfo {author} {\bibfnamefont {C.}~\bibnamefont
  {Balázs}}, \bibinfo {author} {\bibfnamefont {J.}~\bibnamefont {Conrad}},
  \bibinfo {author} {\bibfnamefont {B.}~\bibnamefont {Farmer}}, \bibinfo
  {author} {\bibfnamefont {T.}~\bibnamefont {Jacques}}, \bibinfo {author}
  {\bibfnamefont {T.}~\bibnamefont {Li}}, \bibinfo {author} {\bibfnamefont
  {M.}~\bibnamefont {Meyer}}, \bibinfo {author} {\bibfnamefont {F.~S.}\
  \bibnamefont {Queiroz}}, \ and\ \bibinfo {author} {\bibfnamefont {M.~A.}\
  \bibnamefont {Sánchez-Conde}},\ }\href {\doibase 10.1103/PhysRevD.96.083002}
  {\bibfield  {journal} {\bibinfo  {journal} {Phys. Rev.}\ }\textbf {\bibinfo
  {volume} {D96}},\ \bibinfo {pages} {083002} (\bibinfo {year} {2017})},\
  \Eprint {http://arxiv.org/abs/1706.01505} {arXiv:1706.01505 [astro-ph.HE]}
  \BibitemShut {NoStop}%
%%CITATION = ARXIV:1706.01505;%%
\end{thebibliography}%

\end{document}